\documentclass[10pt,twocolumn,aps,prl,balancelastpage,longbibliography]{revtex4-1}
\usepackage[latin9]{inputenc}
\usepackage{color}
\usepackage{verbatim}
\usepackage{amsmath}
\usepackage{amssymb}
\usepackage{graphicx}
\usepackage{esint}
\usepackage[unicode=true,pdfusetitle,
 bookmarks=true,bookmarksnumbered=false,bookmarksopen=false,
 breaklinks=false,pdfborder={0 0 1},backref=false,colorlinks=true]{hyperref}
\usepackage{breakurl}
\usepackage{epstopdf}

\makeatletter

\providecommand{\tabularnewline}{\\}

\@ifundefined{textcolor}{}
{%
 \definecolor{BLACK}{gray}{0}
 \definecolor{WHITE}{gray}{1}
 \definecolor{RED}{rgb}{1,0,0}
 \definecolor{GREEN}{rgb}{0,1,0}
 \definecolor{BLUE}{rgb}{0,0,1}
 \definecolor{CYAN}{cmyk}{1,0,0,0}
 \definecolor{MAGENTA}{cmyk}{0,1,0,0}
 \definecolor{YELLOW}{cmyk}{0,0,1,0}
}

\usepackage{subfigure}\usepackage{bm}

\newcommand{\bra}[1]{\langle #1 |}
\newcommand{\ket}[1]{|#1\rangle}

\makeatother

\begin{document}

\title{Unified Topological Response Theory for Gapped and Gapless Free Fermions}

\author{Daniel Bulmash, Pavan Hosur, Shou-Cheng Zhang, Xiao-Liang Qi}

\affiliation{Department of Physics, Stanford University, Stanford, California 94305-4045, USA}

\date{\today}
\begin{abstract}
We derive a scheme for systematically characterizing the responses of
gapped as well as gapless systems of free fermions to electromagnetic
and strain fields starting from a common parent theory. Using the
fact that position operators in the lowest Landau level of a quantum
Hall state are canonically conjugate, we consider a massive Dirac
fermion in $2n$ spatial dimensions under $n$ mutually orthogonal
magnetic fields and reinterpret physical space in the resulting zeroth
Landau level as phase space in $n$ spatial dimensions. The bulk topological
responses of the parent Dirac fermion, given by a Chern-Simons theory,
translate into quantized insulator responses, while its edge anomalies
characterize the response of gapless systems. Moreover, various physically
different responses are seen to be related by the interchange of position
and momentum variables. We derive many well-known responses, and demonstrate
the utility of our theory by predicting spectral flow along dislocations
in Weyl semimetals.
\end{abstract}
\maketitle

\section{Introduction}

Spurred by the discovery of topological insulators, topological phases
have become a vital part of condensed matter physics over the last
decade\cite{QiSCZhangTIReview,HasanKaneReview,QiSCZhangTITSCReview,HasanMooreReview}.
Even in the absence of interactions, a wide variety of gapped topological
phases of fermions are now known, ranging from the quantum Hall\cite{HaldaneQHE88,TKNN}
and the quantum spin Hall\cite{KaneMeleQSH,KaneMeleZ2,RoyQSH3D,FuKaneMeleTI3D,Bernevig2006,BernevigZhangHgTe,Roy2DZ2}
insulators among insulators to the chiral $p$-wave superconductor\cite{ReadGreenP+ipFQHE00,KallinSrRuOreview}
and the $B$ phase of Helium-3\cite{OsheroffHe3SF,LeggettHe3SF} among
superconducting phases. All these phases share some common features:
as long as certain symmetry conditions are upheld, they have a bulk
band structure that cannot be deformed into that of an atomic insulator
-- a \emph{trivial} insulator by definition -- without closing the
band gap along the way. Moreover, they all have robust surface states
that mediate unusual transport immune to symmetry-respecting disorder. This leads one to wonder whether all gapped phases of free fermions can be unified within a common mathematical framework.

Two different approaches have been developed to provide unified characterization of gapped phases of free fermions. In the topological band theory approach \cite{KitaevClassification,Ryu2010,SFRLClassification}, homotopy theory and $K$-theory are applied to classify free fermion Hamiltonians in a given spatial dimension and symmetry class. The topological band theory provides a complete topological classification of free fermion gapped states in all dimensions and all the $10$ Altland-Zirnbauer symmetry classes\cite{altland1997}. However, it does not directly describe physical properties of the topological states. In comparison, the topological response theory approach\cite{QiHughesZhangTFT,EssinMPAxion,wang2011,ryu2012,qi2013}  describes topological phases by topological terms in their response to external gauge fields and gravitational fields. The advantage of this approach is that the topological phases are characterized by physically observable topological effects, so that the robustness of the topological phase is explicit and more general than in the topological band theory. Since it is insensitive to details of the microscopic Hamiltonian, a response theory based classification
scheme can be further extended to strongly interacting systems\cite{WangInteractingTI}.

Recently, the advent of Weyl semimetals (WSMs) has triggered interest
in gapless topological phases of free fermions\cite{PyrochloreWeyl,WeylMultiLayer,BurkovNodalSemimetal,TurnerTopPhases,VafekDiracReview,KrempaWeyl,HosurWeylReview}.
These phases are topological in the sense that they cannot be gapped
out perturbatively as long as momentum and charge are conserved. In
this regard, ordinary metals are also topological since their Fermi
surfaces are robust in the absence of instabilities towards density
waves or superconductivity. Additionally, gapless topological phases
may have non-trivial surface states such as Fermi arcs\cite{PyrochloreWeyl,HosurFermiArcs,PotterFermiArcOsc,HaldaneFermiArc}
and\emph{ }flat bands\cite{VolovikFlatBands}. Teo and Kane\cite{TeoKaneTopologicalDefects}
applied homotopy arguments to classify topological defects such as
vortices and dislocations in\emph{ }gapped phases; Matsuura \emph{et.
al.}\cite{Matsuura2013}\emph{ }used an analogous prescription to
classify gapless phases by observing that gapless regions in momentum
space such as Fermi surfaces and Dirac nodes can be viewed as topological
defects in momentum space in a gapped system. Thus, a common mathematical
formalism to describe the Bloch Hamiltonians of gapless phases was
derived.

Unlike their gapped counterparts, however, it is not clear whether
the response theories of gapless phases are amenable to a unified
description. For gapped systems, the path from the Hamiltonian
to the response theory is conceptually straightforward: the fermions
are coupled to gauge fields and integrated out to get the low energy
effective field theory, which describes the topological response properties.
In contrast, the low energy theory of gapless phases contains fermions
as well as gauge fields, and is distinct from the response theory
which contains only gauge fields. Thus, it is not obvious how the
topological properties of the Hamiltonian affect the response.  The
response depends on system details in general and therefore, recognizing
its universal features and then unifying the responses of various
gapless phases is a non-trivial task. A few cases of topological response properties of gapless fermions have been studied. One example is the intrinsic anomalous Hall effect of a two-dimensional Fermi gas\cite{karplus1954,HaldaneAHE,jungwirth2002,XiaoBerryReview,nagaosa2010}. A generalization of this effect in three-dimensional doped topological insulators has been discussed\cite{barkeshli2011}. Another example is the topological response of Weyl semimetals, which has been described in the form of the axial anomaly\cite{adler1969,bell1969,NielsenFermionDoubling1,NielsenFermionDoubling2,NielsenABJ,ZyuninBurkovWeylTheta,ChenAxionResponse,AjiABJAnomaly,QiWeylAnomaly,HosurWeylReview,BasarTriangleAnomaly,SonSpivakWeylAnomaly,LandsteinerAnomaly,GoswamiFieldTheory,GrushinWeyl}. This refers to the apparent charge conservation violation that occurs for each Weyl fermion branch in the presence of parallel electric and magnetic fields, although the net charge of the system must still be conserved. Recently, these ideas were generalized to find the topological responses of point Fermi surfaces in arbitrary dimensions\cite{HughesTopoSemimetal}. The DC conductivity of metals has also been proposed to be related to a phase space topological quantity \cite{HetenyiDrude}. However, a general theory that describes the topological properties of gapless fermions in a unified framework has not been developed yet. 

In this work, we achieve the above goals for free fermions: we show that gapless systems have universal features, independent of system details, and derive
a unified description of their response. Remarkably, this description
also captures the response of gapped systems. In particular, the response
of gapped phases arises from the bulk response of a certain parent
topological phase, while the universal features of the gapless phases
correspond to its edge anomalies. We elucidate this idea below.

The backbone of our construction is a mapping from $n$-dimensional gapped \emph{or} gapless systems to a \emph{gapped} quantum Hall (QH) system which lives in $2n$-dimensional phase space. Such a phase space system has both bulk responses, given by a $2n$-dimensional Chern-Simons (CS) theory, and boundary (axial) anomalies. We identify the bulk responses with topological responses of insulators. However, the key insight that allows us to include gapless systems is to identify a Fermi surface in real space with a phase space boundary in the momentum directions. Likewise, real space excitations near the Fermi surface are identified with the gapless edge excitations in phase space. The universal features of the response of gapless systems are thus the anomalies associated with these phase space edges.

There is an important technical point required in order to bestow the $2n$-dimensional QH system with the interpretation of phase space. Specifically, we must choose the QH system to consist of a
massive Dirac fermion under $n$ uniform magnetic fields of strength
$B_{0}$ in $n$ orthogonal planes, then project to the zeroth Landau level (ZLL) of the total field. In this case, the projected
operators for pairs of dimensions acquire the usual canonical commutation
relations that relate ordinary real and momentum space (up to an overall
factor of $B_{0}$). This allows us to interpret the ZLL of the $2n$-dimensional QH system as phase space for the $n$-dimensional physical space.
We interpret additional perturbations in the phase space gauge fields as physical quantities such as the $n$-dimensional system's electromagnetic (EM)
field, strain field, Berry curvatures, and Hamiltonian. Topological defects, such as monopoles, in the phase space gauge fields allow us to generalize to systems with dislocations and with point Fermi surfaces such as graphene and  Weyl semimetals. These ideas are summarized in Table \ref{table:dictionary}.

\begin{table*}
\begin{centering}
\begin{tabular}{|c|c|}
\hline
\textbf{Phase space}  & \textbf{Real space} \tabularnewline
\hline
Bulk responses  & Quantized insulator responses\tabularnewline
\hline
Anomalies from momentum direction edges  & Gapless response\tabularnewline
\hline
Anomalies from real direction edges  & Real edge anomalies \tabularnewline
\hline
Gauge field strength  & EM field strength/$k$-space Berry curvature/Strain \tabularnewline
\hline
Monopole in gauge field  & Magnetic monopole/Weyl node/dislocation \tabularnewline
\hline
\end{tabular}\caption{Dictionary for interpreting phase space quantities in real space.}

\par\end{centering}

\label{table:dictionary}
\end{table*}

This construction enables us to systematically enumerate all possible
intrinsic, topological responses to electromagnetic and strain fields
in the DC limit in any given dimension. One simply has to write down
the Chern-Simons action in phase space, vary it with respect to each
gauge field, and consider each boundary to obtain all the bulk, boundary
and gapless responses in real space. Following
this procedure, we show carefully that screw dislocations in Weyl semimetals
trap chiral modes which are well-localized around the dislocation
at momentum values away from the Weyl nodes. A related but different effect has been studied previously \cite{jianWSMdisloc}. However, our framework provides a unified and natural description of this effect and other topological effects. 

It is crucial that the $2n$-dimensional system be gapped even in the absence of the background magnetic fields
of strength $B_{0}$. This ensures that its response theory contains
terms depending on $B_{0}$ in addition to fluctuations in the gauge
fields. In $n$ dimensions, we will see that the $B_{0}$-dependent
terms translate into quasi-lower-dimensional responses, such as the
polarization of a system of coupled chains along the chains.  If the
$2n$-dimensional system is gapless in the absence of the background
fields, such responses will be missed by the unified theory.

A caveat is that our construction does not capture responses to spatial,
momental and temporal variations in the field strengths, such as the
gyrotropic effect which is an electric response to a spatially varying
electric field. Note that the regular Maxwell response, given by $j^{\mu}=\partial_{\nu}F^{\nu\mu}$
is a response to a variation in the field strength. Another caveat
is that in phase space dimensions equal to $4$ and above, the Maxwell
term in the action is equally or more relevant than the Chern-Simons
term and hence will, in general, dominate the DC response. However,
our central objective is to demonstrate that there exists a theory
which unifies the responses of gapped and gapless systems, namely,
the phase space Chern-Simons theory.

The rest of this paper is structured as follows. In Section \ref{section:ZLL},
we review the key property of the ZLL which provides the physical
justification for our construction. In Section \ref{section:CSExample}
we explain the interpretation of the gauge fields in our mapping and
give an example illustrating the validity of the CS theory. In Section
\ref{section:diracModel} we write down an explicit model with a CS
response theory and show the precise way in which it behaves as the
phase space response theory of a lower-dimensional model. In Sections
\ref{section:responses} and \ref{section:anomalies} we explain the
responses and anomalies (respectively) that come from the CS theory
in various dimensions, applying our framework to describe spectral
flow in Weyl semimetals with dislocations. Finally, in Section \ref{section:discussion}
we summarize our work and suggest extensions of our theory to more
nontrivial systems.

\section{A Review of the Algebra of the Zeroth Landau Level}

\label{section:ZLL}

One of the key features that we use in the intuition for our approach
is the fact that projecting position operators to the ZLL yields nonzero
commutators between those operators. We now review this fact, for
concreteness as well as for later convenience, for the case of Dirac
electrons in a uniform magnetic field in two spatial dimensions in
Landau gauge. Although we consider the ZLL of Dirac electrons here,
the non-commutativity of position operators is simply a consequence
of minimal coupling and Landau quantization of cyclotron orbits and
hence is true for other dispersions as well as for other Landau levels
for a Dirac dispersion.

Consider a 2D massive Dirac Hamiltonian in a uniform magnetic field
\begin{equation}
H=(p_{x}-eBy)\sigma_{x}+p_{y}\sigma_{y}+m\sigma_{z}
\end{equation}
Here $\sigma_{i}$ are the Pauli matrices. We have set the Fermi velocity
to unity, written the electron charge as $-e$, and chosen the Landau
gauge $\mathbf{A}=-By\mathbf{\hat{x}}$ with $B>0$ for definiteness.
Note that $p_{x}$ commutes with the Hamiltonian, so we may replace
it by its eigenvalue. We can define an annihilation operator $a=(p_{x}-eBy-ip_{y})/\sqrt{eB}$,
which has $[a,a^{\dagger}]=1$, and the Hamiltonian becomes
\begin{equation}
H=\begin{pmatrix}m & \sqrt{2eB}a\\
\sqrt{2eB}a^{\dagger} & -m
\end{pmatrix}
\end{equation}
It is straightforward to show that the eigenstates are labeled by
an eigenvalue $n\geq0$ of the number operator $a^{\dagger}a$, with
dispersion $\pm\sqrt{2eBn+m^{2}}$ for $n\neq0$. For $n=0$, the
eigenvalue is $-m$, the spin state is $\begin{pmatrix}0\\
1
\end{pmatrix}$, and the state is annihilated by $a$. This is the expected result
that the kinetic energy is quenched and the spectrum becomes discrete,
highly degenerate Landau levels.

Let $\ket{k_{x}}$ be the state in the ZLL with $p_{x}$ eigenvalue
$k_{x}$. Then the projection operator to the ZLL is
\begin{equation}
P=\int dk_{x}\frac{L}{2\pi}\ket{k_{x}}\bra{k_{x}}
\end{equation}
with $L$ the system length in the $x$ direction. Writing $y=((p_{x}/\sqrt{eB})-(a+a^{\dagger}))/\sqrt{eB}$,
the projected $y$ operator becomes
\begin{equation}
PyP=\int dk_{x}\frac{L}{2\pi}\frac{k_{x}}{eB}\ket{k_{x}}\bra{k_{x}}
\end{equation}
where we have used the fact the $a$ and $a^{\dagger}$ describe inter-Landau
level processes and thus, vanish under projection onto the ZLL. Next,
using the fact that $\ket{k_{x}}$ is an eigenstate of $p_{x}$, we
find
\begin{equation}
PxP=\int dk_{x}\frac{L}{2\pi}i\partial_{k_{x}}\ket{k_{x}}\bra{k_{x}}
\end{equation}
The commutator can then easily be computed to be
\begin{equation}
\left[PxP,PyP\right]=\frac{i}{eB}\label{eq:[x,y]=00003D00003D00003Di}
\end{equation}

Hence if we absorb the factor of $eB$ into $y$, then $PxP$ and
$PyP$ have the correct commutator structure for us to imbue them
with the interpretation of the position and momentum operators, respectively,
of a 1D system. This interpretation is the primary physical motivation
for the construction which follows. As mentioned earlier, other dispersions
will also result in commutation relations similar to Eq. (\ref{eq:[x,y]=00003D00003D00003Di})
and thus imbue $x$ and $y$ with interpretations of position and
momentum of a 1D system. However, a massive Dirac dispersion is ideal
for deriving the unified response theory because it does not miss
any quasi-lower-dimensional responses, as mentioned earlier and detailed
later.

In higher dimensions, the Dirac model is $H=\sum_{i}(p_{i}+eA_{i})\Gamma_{i}$,
where the $\Gamma_{i}$ are anticommuting elements of the Clifford
algebra of $2n$ by $2n$ matrices. If we apply constant magnetic
fields $F_{ij}$ for disjoint pairs $(i,j)$ of coordinates, we can
form an annihilation operator for each such pair. Annihilation operators
from different pairs commute, and the analysis above carries through
so that the position operators within each pair no longer
commute after projection.

\section{Phase Space Chern-Simons Theory}

\label{section:CSExample}

The key idea of our construction is to represent a (possibly gapless)
$n$-dimensional system by a gapped $2n$-dimensional phase space
system, specifically a massive Dirac model coupled to a gauge field.
As we just showed, we can interpret a $2n$-dimensional system
as living in phase space by adding background magnetic fields between disjoint pairs of spatial directions and projecting to the ZLL. Moreover, since
the phase space system is gapped, we can immediately write down a
response theory for it, the topological part of which can be proved
to be a CS theory\cite{QiHughesZhangTFT, YaoLeeTIChernSimons, Zhang4DQH,Bernevig4DQH}. Note that in a CS theory,
real and momentum space gauge fields enter the action in similar ways,
analogous to our idea of treating position and momentum on the same
footing in phase space.

Before proceeding, we fix some notation and conventions. We will always
use the Einstein summation convention where repeated indices are summed.
Phase space coordinates will be labelled by $x,y,z$ and $\bar{x},\bar{y},\bar{z}$.
After projection, $x,y,z$ will be interpreted as the corresponding
real space coordinates, while $\bar{x},\bar{y},\bar{z}$ will be interpreted
as momentum space coordinates $k_{x},k_{y},k_{z}$ respectively. In
phase space, we will refer to the $U(1)$ background gauge field which
generates the real space commutator structure as $A_{\mu}$ with its nonzero field strengths being $F_{i \bar{i}} = B_0$ for $i=(x,y,z)$. We denote all other contributions to the gauge field by $a_{\mu}$ and the total gauge field by $\mathcal{A}_{\mu}=A_{\mu}+a_{\mu}$.
Likewise, we write $f_{\mu \nu}$ and $\mathcal{F}_{\mu\nu} = F_{\mu \nu} + f_{\mu \nu}$ for the non-background and total field strengths respectively.
The (non-Abelian) field strengths are as usual defined by $f_{\mu\nu}=\partial_{\mu}a_{\nu}-\partial_{\nu}a_{\mu}-[a_{\mu},a_{\nu}]$.

We will also abbreviate the CS Lagrangian by $\epsilon a\partial a\equiv\epsilon^{\mu\nu\sigma}a_{\mu}\partial_{\nu}a_{\sigma}$,
where $\epsilon$ is the totally antisymmetric Levi-Civita tensor,
with an analogous abbreviation for higher-dimensional CS terms. Finally,
we set $e=\hbar=1$, and also assume that $\sqrt{B_{0}}\sim1/l_{B}$
is very large compared to all other wavenumbers in the problem.

\subsection{Interpretation of Phase Space Gauge Fields}

\label{subsection:interpretation}

Our prescription is that the non-background contributions $a_{\mu}$
to the phase space gauge field should be interpreted as the Berry
connection for the lower-dimensional system:
\begin{equation}
a_{\mu}^{\alpha\beta}=i\bra{u_{k\alpha}}\partial_{\mu}\ket{u_{k\beta}}
\end{equation}
where $\ket{u_{k\alpha}}$ is the (local) Bloch wavefunction at momentum $k$
for the $\alpha$ band. Here $\partial_{\bar{x},\bar{y},\bar{z}}$
should be interpreted as $B_{0}\partial_{k_{x},k_{y},k_{z}}$. We can think of the physical EM vector potential as a Berry connection, which means that it is included in the real-space components of $a$.

As such, we will often use the following heuristic interpretations
in order to more clearly see the physics: $-a_{t}$ is the lower-dimensional
band Hamiltonian plus the physical EM scalar potential, $a_{x,y,z}$ is the
physical EM vector potential, and $a_{\bar{x},\bar{y},\bar{z}}$ is the momentum-space
Berry connection. The field strengths which do not mix real and momentum
space then have natural interpretations as the physical EM field strengths
and Berry curvatures.

The physical interpretation of ``mixed'' field strengths such as $f_{x\bar{y}}$ (in 4 or higher
dimensional phase space) is less obvious. Here we present two ways
to think about them. First consider a gauge where $\partial_{\bar{y}}a_{x}=0$.
We find that
\begin{equation}
\int d\bar{y}f_{x\bar{y}}=\partial_{x}\int d\bar{y}a_{\bar{y}}=2\pi\partial_{x}P_{y}
\end{equation}
where $P_{y}$ is the one-dimensional polarization of the system\cite{Vanderbilt1993}.
A spatially varying polarization can be thought of as strain of the
electron wavefunction, which can come from either mechanical strain
or some other spatial variation of the parameters entering the band
structure.

To make the connection of $f_{x\bar{y}}$ to mechanical strain more
explicit, we change the gauge to set $\partial_{x}a_{\bar{y}}=0$. An
intuitive way to think about a nonzero $f_{x\bar{y}}$ in this gauge
is in terms of dislocations. In particular, adiabatically moving a
particle around a real space dislocation leads to a translation, but
if the particle can locally be treated as a Bloch wave, then that
translation is equivalent to the accumulation of a phase. This (Berry)
phase is equal to $\mathbf{k}\cdot\mathbf{b}$, with $\mathbf{b}$
the Burgers vector of the dislocation. In particular, this is a \emph{momentum-dependent}
Berry phase resulting from adiabatic motion in real space. Hence $f_{x\bar{y}}$
is nonzero. It can be shown explicitly\cite{XiaoBerryReview} in the
perturbative regime that strain typically leads to such a Berry phase.

\subsection{Example: 2D Phase Space}

\label{section:2DExample}

\begin{figure}
\subfigure[ ]{ \includegraphics[width=4cm]{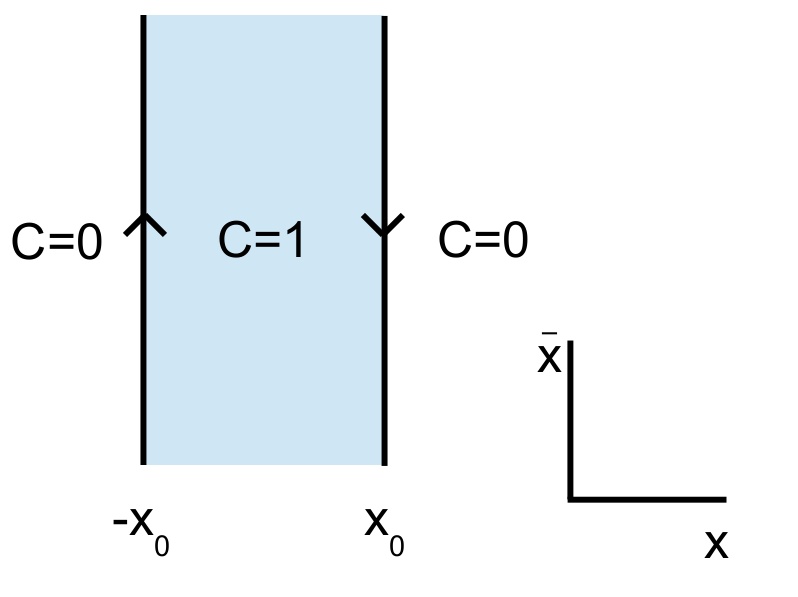} \label{fig:realEdge}
} \subfigure[ ]{ \includegraphics[width=4cm]{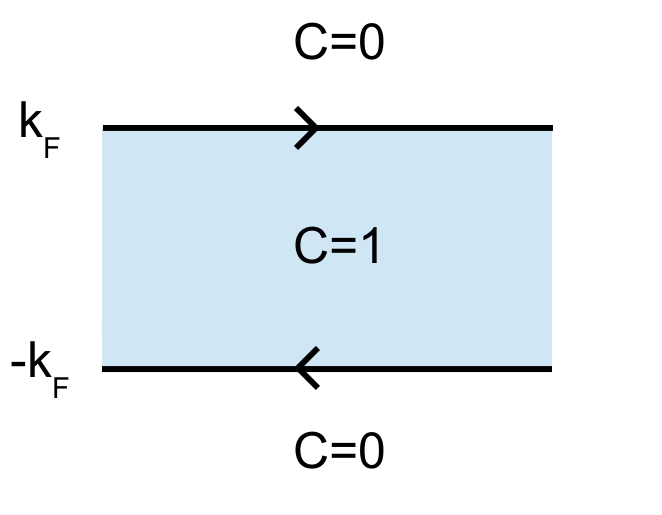} \label{fig:FSEdge}
} \caption{Phase space realization of (a) real-space edges (b) Fermi points for
a 1D real space system. Arrows indicate the direction of the edge
modes.}
\end{figure}

We first consider the case where our real space theory consists of
a single filled band living in 1D and that a 2D CS response term in
phase space with a background field describes the expected responses.
We will, for simplicity, only consider Abelian physics in this example.
Consider the CS action
\begin{equation}
S_{CS}=\frac{1}{4\pi}\int dtd^{2}xC(\bar{x},x)\epsilon\mathcal{A}\partial\mathcal{A}\label{eqn:ChernSimons2D}
\end{equation}
(In Section \ref{section:diracModel}, we will show in an explicit model how Eq. \ref{eqn:ChernSimons2D}, with this (quantized) coefficient, appears, but for now we simply assume that it is the relevant response theory.) Here $C(\bar{x},x)$ accounts for the filling at different points;
for example, if the system occupies $x>0$, then $C(\bar{x},x)$ will
be proportional to $\Theta(x)$ with $\Theta$ the Heaviside step
function, as shown in Fig. \ref{fig:realEdge}. Likewise, if the system
has a Fermi momentum $k_{F}$, then $C(\bar{x},x)$ will be proportional
to $\left(\Theta(\bar{x}+k_{F}/B_{0})-\Theta(\bar{x}-k_{F}/B_{0})\right)$,
as shown in Fig. \ref{fig:FSEdge}.

Let us assume that there are no edges so that $C=1$ everywhere. Then,
the responses for this action, given by $j^{\mu}=\delta S/\delta\mathcal{A}_{\mu}$,
are
\begin{equation}
j_{2D}^{\mu}=\frac{1}{2\pi}\epsilon^{\mu\nu\sigma}\mathcal{F}_{\nu\sigma}
\end{equation}
where $\mathcal{F}_{\nu\sigma}=\partial_{\nu}\mathcal{A}_{\sigma}-\partial_{\sigma}\mathcal{A}_{\nu}$
is the field strength tensor corresponding to $\mathcal{A}$. Let us consider
each component, assuming for conciseness that the background field
is in Landau gauge $A_{\bar{x}}=B_{0}x$.

First we examine the real-space response $j_{2D}^{x}=\mathcal{F}_{\bar{x}t}/2\pi$.
This current in general depends on $\bar{x}$, which we interpret
as $k_{x}/B_{0}$; the observable current should then be given by
integrating the 2D current with respect to $\bar{x}$, as the real-space
current has contributions from all occupied momenta. The resulting
1D current is
\begin{equation}
j_{1D}^{x}=\frac{1}{2\pi}\left(\int d\bar{x}\partial_{\bar{x}}a_{t}-\partial_{t}\int d\bar{x}a_{\bar{x}}\right)\label{eq:jx1D}
\end{equation}
Interpreting the $\bar{x}$-dependent part of $a_{t}$ as the dispersion,
the first integral generically gives zero. The second integral is,
for a gapped system, exactly the time derivative of the polarization $P_{x}=\frac{1}{2\pi}\int a_{\bar{x}}\mathrm{d}\bar{x}$,
which is the expected 1D real-space current response $j_{1D}^{x}=-\partial_{t}P_{x}$.
Similarly, the $k$-space response is $j_{2D}^{\bar{x}}=\mathcal{F}_{tx}/2\pi$.
Interpreting $j^{\bar{x}}_{2D}$ as $dk/dt$, we recover the real space semiclassical
equation of motion $dk/dt=E/2\pi$ with $E$ the electric field.

Finally, the charge response is given by
\begin{equation}
\rho_{1D}=\frac{1}{2\pi}\int\mathrm{d}\bar{x}\left(F_{x\bar{x}}+\partial_{x}a_{\bar{x}}-\partial_{\bar{x}}a_{x}\right)\label{eq:rho1D}
\end{equation}
In units $B_{0}=1$, the first term simply gives the total charge
in the occupied band, which can be thought of a quasi-0D response.

The second term of Eq. (\ref{eq:rho1D}), in a gauge where $a_{x}=0$,
becomes $\partial_{x}P_{x}$ for a gapped system. This is again intuitive;
if, say, the system is strained, then the polarization and hence the
charge density will change accordingly.

If we now impose a pair of edges at $\bar{x}=\pm k_{F}$, two things
happen. Firstly, the 1D system lies between a pair of momentum points
$\pm k_{F}$, so the integrals in (\ref{eq:jx1D}) and (\ref{eq:rho1D})
run from $-k_{F}$ to $+k_{F}$ instead of the full Brillouin zone.
Consequently, the background charge becomes $\rho_{1D}^{bg}=\frac{1}{2\pi}\intop_{-k_{F}}^{k_{F}}F_{x\bar{x}}=k_{F}/\pi$,
as expected, while the terms proportional to $f_{x\bar{x}}$ cease
to have a simple interpretation as the polarization but can be non-zero
nonetheless. Secondly, the 2D system develops a chiral anomaly at
the edge, given by
\begin{equation}
\partial_{t}\rho_{2D}+\partial_{x}j_{2D}^{x}=\frac{1}{2\pi}\mathcal{F}_{tx}\partial_{\bar{x}}f(\bar{x})=\frac{E+\partial_{x}\varepsilon}{2\pi}\partial_{\bar{x}}C(\bar{x})\label{eqn:FSanomaly}
\end{equation}
where $C(\bar{x})=\Theta(\bar{x}+k_{F})-\Theta(\bar{x}-k_{F})$. Integrating
over 1D real space under a constant electric field and translational
invariance yields
\begin{equation}
\partial_{t}\int dx\rho_{2D}=\frac{L}{2\pi}\left(\delta(\bar{x}+k_{F})-\delta(\bar{x}-k_{F})\right)E
\end{equation}
where $L$ is the length of the system and $\delta$ is the Dirac
delta function. This is precisely the chiral anomaly in the 1D system:
the electric field tilts the 1D Fermi surface, effectively converting
right-moving charge in the vicinity of one Fermi point into left-moving
charge near the other. Thus, we have derived a property of a gapless
1D band structure from the edge anomaly of the parent 2D QH system.

Notice also that integration of (\ref{eqn:FSanomaly}) over momentum
space leads to
\begin{equation}
\partial_{t}\rho_{1D}+\partial_{x}j_{1D}^{x}=0
\end{equation}
which correctly tells us that there is no anomaly in the total charge.
The precise value of $\rho_{1D}$ and $j_{1D}^{x}$ depends on system
details; therefore, calculating them in our formalism would require
knowledge of non-universal properties of the 2D QH edge, such as the
velocity of the chiral modes. However, we have shown here that they
still have universal properties that reflect the universal properties
of a higher dimensional topological state.

A different type of anomaly occurs when the system has real-space
edges and a filled band. In this case, the anomaly equation in 2D
is
\begin{equation}
\partial_{t}\rho_{2D}+\partial_{\bar{x}}j_{2D}^{\bar{x}}=-\frac{1}{2\pi}\mathcal{F}_{t\bar{x}}\partial_{x}C(x)
\end{equation}
Integrating the above in $\bar{x}$ yields
\begin{equation}
\partial_{t}\rho_{1d}=(\left(\delta(x-x_{0})-\delta(x+x_{0})\right)\partial_{t}P(x)
\end{equation}
This is the known result\cite{ThoulessChargePump} that charge can be adiabatically pumped from
one edge of the system to the other via a time-dependent local polarization.

We thus see that the standard responses, including anomalies, that
we expect in a 1D theory are retrieved from the 2D CS theory. However,
detail-dependent edge responses are described in our theory only by
the anomaly (or lack thereof) that they create. We expect the same
procedure to generalize to higher dimensions, and we will show that
the expected topological responses appear in Section \ref{section:responses}.

\section{Explicit Model}

\label{section:diracModel}

\begin{figure}
\centering{}\includegraphics[width=0.95\columnwidth]{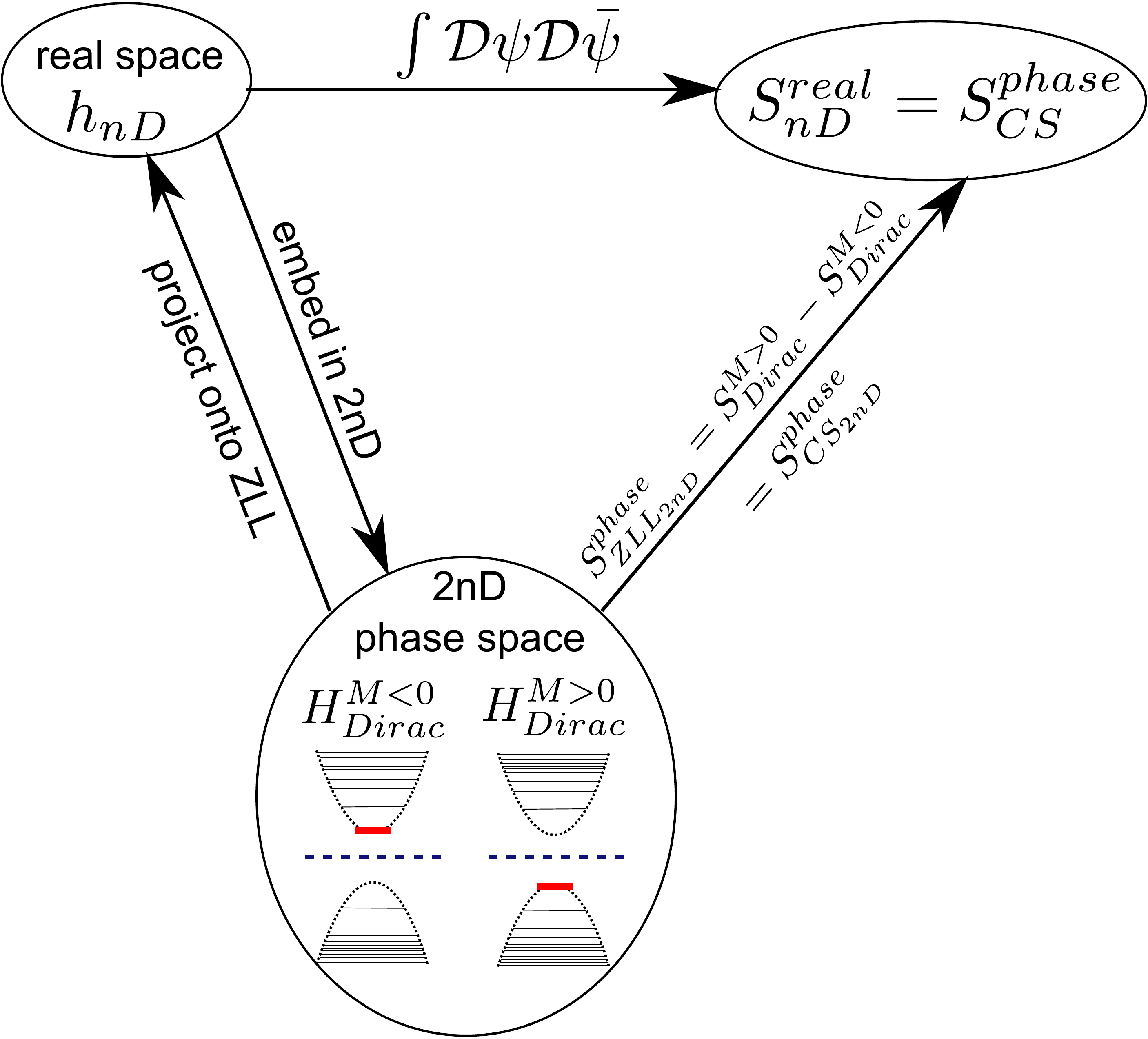}
\caption{Logical flow of the derivation in Section \ref{section:diracModel}.
An $nD$ Hamiltonian $h_{nD}$ can be obtained by projecting a $2nD$
massive Dirac Hamiltonian $H_{Dirac}^M$ in a magnetic field onto the ZLL, denoted
by the thick red bar. For a Fermi level in the Dirac mass gap, the
$M>0$ and $M<0$ ground states differ only in the occupation of the
ZLL, while their response theories $S_{Dirac}^M$ differ by the CS-term in $2nD$.
Thus, the response theory of the ZLL $S_{ZLL_{2nD}}$, which is the real space response theory $S_{nD}^{real}$ of $h_{nD}$, is the phase space $CS$-theory $S_{CS}^{phase}$.\label{fig:derivationLogic}}
\end{figure}

In this section, we elucidate the precise way in which a ZLL behaves
as the phase space of a system in half the number of spatial dimensions.
In particular, we explain why the response theory of the lower dimensional
system should be of CS form in phase space and describe the physical
meaning of projecting onto the ZLL. We also answer the question of
when a CS theory in $2nD$ can be interpreted as a phase space response
theory in $nD$.

To begin, consider a massive Dirac Hamiltonian in $2n$ dimensions
coupled to the gauge field $\mathbf{\mathcal{A}}$ defined in
Sec. \ref{section:CSExample}:
\begin{equation}
H_{2nD}=\sum_{i=1}^{n}\left[\Gamma_{i}(p_{i}-\mathcal{A}_{i})+\Gamma_{\bar{i}}(p_{\bar{i}}-\mathcal{A}_{\bar{i}})\right]+\Gamma_{0}M
\label{eqn:2nDirac}
\end{equation}
$\mathbf{\mathcal{A}}$ corresponds to large constant background
fields $B_{0}$ in $n$ orthogonal planes plus small fluctuations;
thus, $F_{i\bar{i}}=B_{0}\gg f_{ij},f_{i\bar{j}},f_{\bar{i}\bar{j}}$ for all $i,j\in0\dots n$.
The $\Gamma$'s are $2n\times2n$ anticommuting matrices with eigenvalues
$\pm1$ and satisfy $\Gamma_{0}=\prod_{i=1}^{n}\Gamma_{i}\Gamma_{\bar{i}}$.
To zeroth order in $f$, the spectrum of $H_{2nD}$ can be easily
derived by generalizing the calculation of Sec \ref{section:ZLL}; it consists of Landau levels with energies $\pm\sqrt{2kB+M^{2}}$
for positive integers $k$ together with a ZLL state which has energy $-M$ and a spinor wavefunction that has a $\Gamma_{0}$ eigenvalue of $-1$. 

We have two tasks. First, we must isolate the topological response theory of the ZLL of this system, which we expect to be a Chern-Simons theory. Second, we must relate this Hamiltonian, projected onto the ZLL of the total field, to the Hamiltonian of the real space system.

For the first task, note that the ZLL is occupied (unoccupied) in the ground state if $M>0$ ($M<0$),
while the occupation of all the other Landau levels is independent
of the sign of $M$. This should hold for non-zero $f$ as well if
$M\gg\sqrt{B_{0}}$. As a result, the response of the ZLL to $\mathbf{\mathcal{A}}$
is given by the terms in the total response that are \emph{odd} in $M$.
Moreover, it is known that the two signs of $M$ correspond to a topological
and a trivial insulator (which sign corresponds to which phase is determined by the regularization
far away from the Dirac point).
Therefore, the difference between their response theories, which
by definition is the topological part of the effective action, equals
the response of the ZLL. In the absence of vertex corrections, this
is known to be the $2n$-dimensional CS action with coefficient 1 to lowest order in the coupling constant $e$. In short, the response of the ZLL is precisely
the CS action with coefficient 1 in appropriate dimensions under suitable well-controlled
perturbative approximations. We emphasize that this statement is true
even if $H_{2nD}$ is modified at high energies to change the total
($n^{\text{th}}$) Chern numbers of the occupied and unoccupied bands. The
only requirement is that the Chern numbers of the $M>0$ and $M<0$ cases differ by unity; their
actual values are irrelevant for determining the ZLL response.

Next, we recall that $[x_{i},\bar{x}_{i}]=i$ in the ZLL as shown
in Sec \ref{section:ZLL}, so $x_{i}$ and $\bar{x}_{i}$ can be thought
of as a pair of canonically conjugate position and momentum variables.
Therefore, projecting $H_{2nD}$ onto the ZLL gives an $n$-dimensional
system whose response theory is guaranteed to be of CS form in phase
space. In this response theory, the gauge fields in the ``momentum''
directions are to be reinterpreted as momentum space Berry connections.
This flow of logic is depicted in Fig \ref{fig:derivationLogic} (where
we have renamed $H_{2nD}$ as $H_{Dirac}^{M}$ to make the figure
self-contained).

Having shown that the response of the $n$-dimensional system is given
by the phase space CS theory, we turn to our second task and show in detail how the Hamiltonian
in $n$-dimensions is related to $H_{2nD}$. For clarity, we choose
$n=1$, i.e., we demonstrate this in 1D real space with a $U(1)$
gauge field; the procedure generalizes straightforwardly to more dimensions
and to larger gauge groups. To construct the 2D phase space model,
let $\Gamma_{x}$, $\Gamma_{\bar{x}}$, and $\Gamma_{0}$ be the Pauli
matrices $\sigma_{x}$, $\sigma_{y}$, $\sigma_{z}$. (The $\Gamma$
notation is for consistency with the higher-dimensional generalization in Eq. (\ref{eqn:2nDirac}).) The
appropriate 2D Hamiltonian is
\begin{equation}
H_{2D}=(p_{x}-\mathcal{A}_{x})\Gamma_{x}+(p_{\bar{x}}-\mathcal{A}_{\bar{x}})\Gamma_{\bar{x}}+M\Gamma_{0}+\mathcal{A}_{0}\label{eqn:2dH}
\end{equation}

We are projecting onto the ZLL of the total field $\mathcal{A}$,
so we need to make some approximations to make progress. We assume
that the field fluctuations $f$ are much smaller than $B_{0}$. That is, we identify $1/\sqrt{B_{0}}$ with some microscopic length
scale like a lattice constant for the underlying real space system,
and assume that all the gauge field fluctuations are small over that
length scale. If this is true, then we can make the gauge choice that
$\partial_{\mu} a_{\nu} \ll B_0$ for all $\mu, \nu$. In this case, the Hamiltonian of the ZLL
of $\mathcal{A}$ can be computed by considering $a$ to be a perturbation
on the Hamiltonian with $\mathcal{A}=A$. We implement degenerate
perturbation theory as follows.

Let us write
\begin{equation}
H_{2D}=H_{0}+H'
\end{equation}
with
\begin{align}
H_{0} & =(p_{x}-A_{x})\Gamma_{x}+(p_{\bar{x}}-A_{\bar{x}})\Gamma_{\bar{x}}+M\Gamma_{0}-A_{0}\\
H' & =-a_{x}\Gamma_{x}-a_{\bar{x}}\Gamma_{\bar{x}}-a_{0}
\end{align}
Since $A$ is a constant background field of strength $B_{0}$, we
know how to diagonalize $H_{0}$; let $\ket{n,k}$ be the eigenstates
where $n$ labels the LL and $k$ labels a momentum (in Landau gauge).
Using $\left\langle \bar{x}|0,k\right\rangle \propto e^{-B_{0}(\bar{x}-k/B_{0})^{2}/2}(1,0)$,
where the spinor indicates that the ZLL states are polarized in the
basis of $\Gamma_{0}$ eigenstates, and denoting $k_{\pm}=k\pm q/2$,
first order degenerate perturbation theory gives an effective 1D Hamiltonian
as
\begin{align}
\bra{0,k_{-}}H'\ket{0,k_{+}} & \propto-\int dxd\bar{x}e^{iqx}a_{0}(x,\bar{x})e^{-B_{0}(\bar{x}-k/B_{0})^{2}}e^{-q^{2}/4B_{0}}\nonumber \\
 & \propto-a_{0}\left(-i\delta'(q),k/B_{0}\right)\nonumber \\
\implies h_{1d}(k) & \equiv-a_{0}\left(i\partial_{k},k/B_{0}\right)\label{eq:firstorderpert}
\end{align}
Thus, the desired 1D Hamiltonian $h_{1d}(x,k)$ can easily be obtained
by choosing $a_{0}(x,\bar{x})=-h_{1d}(x,B_{0}\bar{x})$. Since the
ZLL is spin-polarized, the dependence on $a_{x}$ and $a_{\bar{x}}$
disappears from (\ref{eq:firstorderpert}); these fields only appear
at second order in perturbation theory. Degenerate perturbation theory
tells us that, if $P$ is the projector onto the degenerate subspace,
then the second order correction to the energy is given by the eigenvalues
of\begin{widetext}
\begin{equation}
\bra{\psi_{i}}H_{2D}\ket{\psi_{j}}=\bra{0,k_{i}}\left(H_{0}+H'+(H'-PH'P)(H_{0}-E_{0})^{-1}(H'-PH'P)\right)\ket{0,k_{j}}\equiv\bra{0,k_{i}}H_{2D}+H^{(2)}\ket{0,k_{j}}\label{eqn:pertTheory}
\end{equation}
where
\begin{equation}
\ket{\psi_{i}}=\ket{0,n_{i}}+\sum_{n>0,l}\ket{n,k_{l}}\bra{n,k_{l}}(H_{0}-E_{0})^{-1}(H'-PHP)\ket{n=0,k_{i}}
\end{equation}
\end{widetext} is a basis for the perturbed ZLL wavefunctions up
to first order in $H'$. In particular, the unitary transformation
$U$ which takes $H_{2D}$ to $H_{2D}+H^{(2)}$, to second order in
$H'$, is the one which takes $\ket{0,k_{i}}$ to a state living in
the ZLL of the full Hamiltonian, to first order in $H'$.

Therefore, if we find this unitary transformation and then perform
the projection in the ZLL of the background field, we still get our
desired projected Hamiltonian. We write $U=\exp(iS)$ with $S$ Hermitian,
and expand $S=S_{1}+S_{2}+...$ where the subscripts indicate an expansion
in orders of $H'$ (by inspection $S$ can be chosen to have no zeroth
order term). Then we can match, order by order, terms in $e^{iS}H_{2D}e^{-iS}$
with those in $H_{2D}+H^{(2)}$ to find the conditions
\begin{align}
[H_{0},S_{1}] & =0\label{eqn:firstOrderCondition}\\
[H_{0},S_{2}] & =iH^{(2)}
\end{align}
We do not claim to be able to demonstrate explicitly a unitary transformation
which obeys the second of these conditions, as computing $H^{(2)}$
is highly nontrivial. However, we will proceed first by exhibiting
an ansatz for $U$, then showing that the projection onto the ZLL
of $A$ yields the correct Hamiltonian in real space, and finally
giving the physical motivation for the ansatz.

Let us start in the gauge $A_{x}=-B_{0}\bar{x}$, $A_{\bar{x}}=A_{t}=0$.
Then let
\begin{equation}
U=\left(e^{ia_{\bar{x}}p_{x}ds/B_{0}}\right)^{N}e^{-iB_{0}x\bar{x}}\left(e^{-ia_{x}p_{\bar{x}}ds/B_{0}}\right)^{N}e^{iB_{0}x\bar{x}}
\end{equation}
where $ds$ is an infinitesimal parameter and $N\rightarrow\infty$
such that $Nds=1$. This transformation is a gauge transformation,
followed by a translation of $\bar{x}$ by $-a_{x}$, followed by
the reverse gauge transformation, followed by a translation of $x$
by $a_{\bar{x}}$.

By inspection $U$ commutes with $H_{0}$, so Eq. (\ref{eqn:firstOrderCondition})
is satisfied. To second order in $H'$, we now have \begin{widetext}
\begin{equation}
UH_{2D}U^{\dagger}\approx H_{0}-a_{x}\left(x+\frac{a_{\bar{x}}}{B_{0}},\bar{x}-\frac{a_{x}}{B_{0}}\right)\Gamma_{x}-a_{\bar{x}}\left(x+\frac{a_{\bar{x}}}{B_{0}},\bar{x}-\frac{a_{x}}{B_{0}}\right)\Gamma_{\bar{x}}-a_{0}\left(x+\frac{a_{\bar{x}}}{B_{0}},\bar{x}-\frac{a_{x}}{B_{0}}\right)
\end{equation}
\end{widetext} where $a_{i}$ appearing without explicit functional
dependence means $a_{i}(x,\bar{x})$. The terms that we have neglected are ``double-nestings" of $a/B_0$; our aforementioned approximation that $a$ is slowly varying (which was a gauge choice possible when the corresponding field strengths were weak) allows us to write
\begin{equation}
a_0\left(x+\frac{a_{\bar{x}}}{B_{0}},\bar{x}-\frac{a_{x}\left(x+\frac{a_{\bar{x}}}{B_0},\bar{x}\right)}{B_{0}}\right)\approx a_0\left(x+\frac{a_{\bar{x}}}{B_{0}},\bar{x}-\frac{a_{x}}{B_{0}}\right) 
\end{equation}

Let us now perform the projection
on the ZLL of the background field. As before, everything projects
to zero except for the $a_{0}$ term and the mass term of $H_{0}$.
The latter just projects to a constant which we can absorb by a shift
of $a_{0}$. However, we now obtain a different 1D Hamiltonian $h_{1d}'(x,\bar{x})=-a_{0}(x+a_{\bar{x}},\bar{x}-a_{x}))$,
which is simply $h_{1d}(x,\bar{x})$ with minimal coupling to the
gauge fields $a_{\bar{x}}$ and $a_{x}$, respectively ($B_{0}$ set
to unity for convenience). We therefore have correctly retrieved the
full 1D Hamiltonian from a projection to the ZLL, as the functional
form of the projected Hamiltonian is correct if we imbue $x$ and
$\bar{x}$ with the interpretations of a parameter tracking a locally
periodic Hamiltonian in space and Bloch momentum respectively.

A major question remains: why, physically, should this choice of $U$
be the correct one? First of all, the projected Hamiltonian, if it
is to describe a real system, must be gauge invariant. Hence the gauge
fields should be minimally coupled, and $U$ indeed accomplishes this
goal.

A more fundamental reason, though, is the following. Consider $H_{2D}$
in some local region over which $a$ is approximately constant, and
for convenience choose a gauge in which $a_{\bar{x}}$ is zero. In
this region, $a_{x}$ functions as a constant shift of the momentum
$p_{x}$ which dictates, in the ZLL of $A$, the wavefunction center
in $\bar{x}$. Hence we should, roughly speaking, identify the (local)
eigenvalue of $p_{x}-a_{x}$ with $\bar{x}$. In the original basis,
then, the variable canonically conjugate to $x$ is identified in
the ZLL with $\bar{x}+a_{x}$. If we are to interpret the commutator
of the projected $x$ and $\bar{x}$ operators in phase space as being
the canonical commutation relation of $x$ and $p$ in real space,
then we need to shift $\bar{x}$ by $-a_{x}$ in order to do so. By
a similar argument in the gauge where $a_{x}=0$, we should shift
$x$ by $a_{\bar{x}}$ to identify $x$ with $p_{\bar{x}}$ in the
ZLL.

Having derived the real space Hamiltonian from an ansatz for the solution
to the phase space one, we now comment on a few details.

First notice that this derivation generalizes easily to higher dimensions,
as the background field only couples $x$ to $\bar{x}$, $y$ to $\bar{y}$,
etc. The primary difference is that in $2n$-dimensional phase space,
the $\Gamma$ matrices must be anticommuting elements of the Clifford
algebra of $2n$ by $2n$ matrices with diag$(\Gamma_{i})=0$ for
$i\neq0$.

We next comment on gauge invariance. It may appear that there is extra
gauge invariance in the phase space theory; in particular, it may
seem strange that the Berry connections $a_{\bar{x}}$ can be gauge
transformed into real space gauge fields $a_{x}$ and vice-versa.
We claim that this is simply a reflection of the usual gauge invariance
in the lower-dimensional Hamiltonian. To see this, consider a unitary
operator $U=\exp(if(x,\bar{x}))$ which implements the gauge transformation
$a_{\mu}\rightarrow a_{\mu}+\partial_{\mu}f$, and let the ZLL wavefunctions
be $\ket{n}$ for some set of labels $n$. Since $U$ is a gauge transformation
in the phase-space system, it must commute with the projection operator
$P$ (as $U$ must take states in a given LL to the same LL). Hence
we can project $U$ to get its action on the projected Hamiltonian;
by the same argument we used for projecting the Hamiltonian, we must
have $PUP=\exp(if(x,k))$. To understand the meaning of this operator,
recall that locally, any state can be labeled as a Bloch wavefunction
$\ket{k;x}$ at momentum $k$ for a local Hamiltonian at $x$. Therefore,
a gauge transformation in the higher-dimensional system is equivalent
to a spatially dependent $U(1)$ gauge transformation on the eigenstates
$\ket{k;x}$ of the local Hamiltonian parametrized by $x$.

Finally, after seeing the derivation, we may answer the following
question: when can a CS theory in $2n$D be interpreted as the phase
space response theory of a system in $n$D? The key physical requirement
in our derivation was that the total field in the $2n$D system could
be separated into two parts: a uniform background field, which sets
some length scale, and another portion which varies slowly on that
length scale. When this condition holds, the CS theory may be interpreted
as a phase space theory for some lower-dimensional system.

\section{Enumeration of Bulk Responses}

\label{section:responses}

Having shown that the phase space CS theory is the correct unified
theory, we now systematically enumerate all the bulk responses of
the CS theory for each possible dimensionality of phase space, and
interpret them in real space. To avoid cluttering the notation, we
set $B_{0}=1$.

The 2D responses were discussed in Section \ref{section:2DExample}.
There we showed that the real space current density is the rate of
change of polarization, while the $k$-space current density reflects
the expected relation $dk/dt\sim E$ with $E$ the electric field.
The charge density response is just the band filling corrected for
strain-induced changes in the lattice constant.

We summarize the 2D responses in Table \ref{table:2Dresponses}.

\begin{table*}
\begin{tabular}{|c|c|}
\hline
\textbf{Current component}  & \textbf{Response}\tabularnewline
\hline
Real space  & Change in polarization\tabularnewline
\hline
$k$-space  & Electric force\tabularnewline
\hline
Charge density  & Band filling \tabularnewline
\hline
\end{tabular}\caption{Summary of 2D phase space responses.}

\label{table:2Dresponses}
\end{table*}

\subsection{4D Phase Space}

The action is given by
\begin{equation}
S=\frac{1}{24\pi^{2}}\int dtd^{4}x\text{tr}\left[C(x,\bar{x})\epsilon\mathcal{A}\partial\mathcal{A}\partial\mathcal{A}+...\right]
\end{equation}
where $+...$ indicates the non-Abelian terms. We set $C = 1$ uniformly to look at bulk responses.

\noindent \textbf{Spatial components:}

A priori there is no difference between the spatial directions $x$
and $y$, so we focus on the $x$ responses.

\begin{equation}
j_{2D}^{x}=\frac{1}{4\pi^{2}}\int d^{2}\bar{x}\text{tr}\left[\mathcal{F}_{y\bar{y}}\mathcal{F}_{t\bar{x}}+\mathcal{F}_{\bar{x}\bar{y}}\mathcal{F}_{ty}+\mathcal{F}_{y\bar{x}}\mathcal{F}_{\bar{y}t}\right]
\end{equation}

\begin{figure}
\subfigure[ ]{\includegraphics[width=2.8cm]{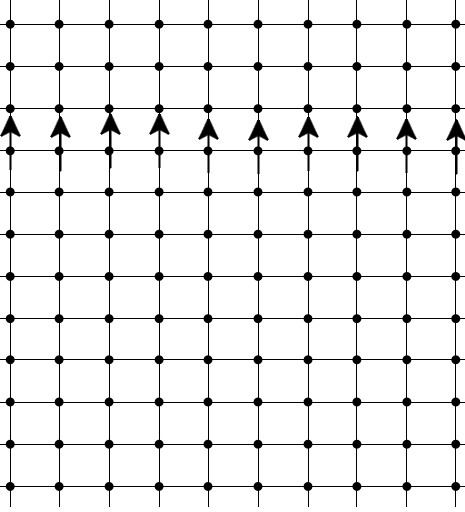} \label{fig:unstrainedPolarization}}
\subfigure[ ]{\includegraphics[width=5.5cm]{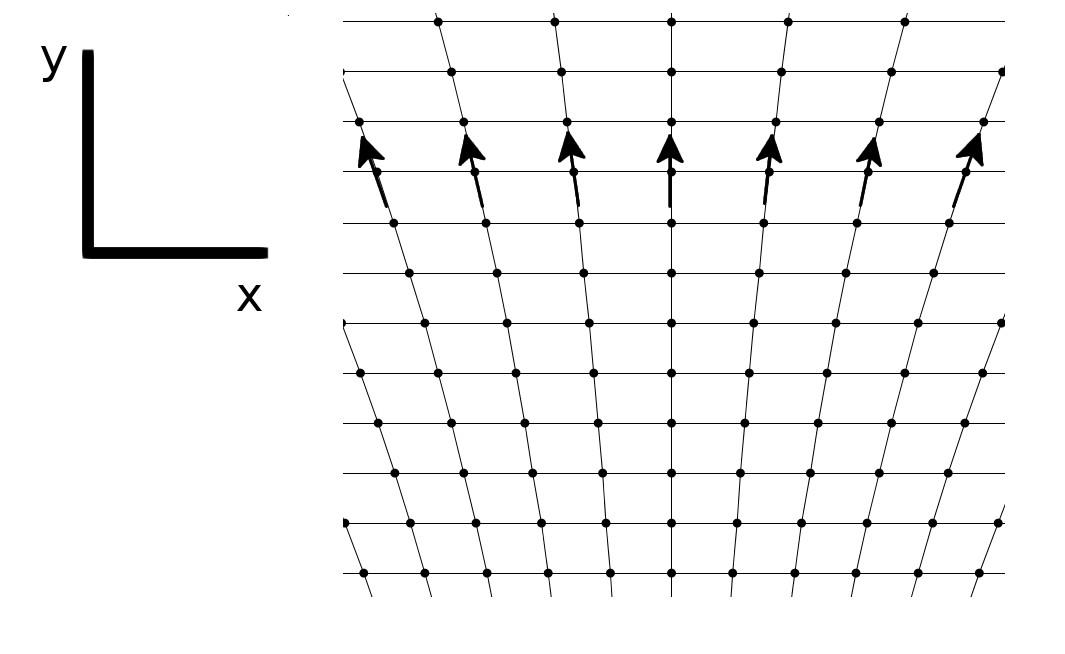} \label{fig:strainedPolarization}}
\caption{Cartoon of the polarization response of an (a) unstrained square lattice (b) sheared square lattice. In (b), $F_{y\bar{x}} \neq 0$ because motion along the $y$ lattice vector leads to translation in the unstrained $x$ direction. The directed lines are the flow of charge due to a positive $\partial_t P_y$; the polarization is measured along the lattice vector, which is deformed in (b) due to strain. Only (b) has a nonzero current in the $x$ direction in (b), which is due to this deformation.}
\end{figure}

The first term includes the background field; setting $\mathcal{F}_{y\bar{y}}=F_{y\bar{y}}=B_{0}$
turns this into $j_{2D}^{x}=\frac{1}{2\pi}\int d\bar{x}\text{tr}\left[\mathcal{F}_{t\bar{x}}\right]=\partial_{t}P_{x}$
with $P_{x}$ the polarization in the $x$ direction. This is the
same response that appears in 1D, and is illustrated in Fig. \ref{fig:unstrainedPolarization}. The second term is the anomalous
Hall response; in the simple case where $\mathcal{F}_{ty}$ is simply
an electric field, this term gives a current $j^{x}=E_{y}\int d^{2}\bar{x}\text{tr}[\mathcal{F}_{\bar{x}\bar{y}}]/4\pi^{2}=E_{y}C_{1}/2\pi$
with $C_{1}$ the first Chern number of the occupied bands. This formula also applies to systems with open boundary in the $\bar{x},\bar{y}$ directions, in which case $C_1$ is not quantized but still determines the intrinsic Hall conductivity of the two-dimensional Fermi liquid\cite{karplus1954,HaldaneAHE,jungwirth2002,XiaoBerryReview,nagaosa2010}.

The third term, illustrated in Fig. \ref{fig:strainedPolarization}, says the following. Suppose that there is a change
in time of the polarization in the $y$-direction, i.e. $\mathcal{F}_{\bar{y}t}\neq0$,
without any strain in the system. If we now add shear in the system,
i.e. have $\mathcal{F}_{y\bar{x}}\neq0$, then some of that polarization
change becomes a current along the $x$ direction as defined before
adding strain. 

\noindent \textbf{$k$-space components}:

The $\bar{x}$-direction responses are:
\begin{equation}
j_{4D}^{\bar{x}}=\frac{1}{4\pi^{2}}\text{tr}\left[\mathcal{F}_{tx}\mathcal{F}_{y\bar{y}}+\mathcal{F}_{ty}\mathcal{F}_{\bar{y}x}+\mathcal{F}_{t\bar{y}}\mathcal{F}_{xy}\right]
\end{equation}
The first term is quasi-1D, saying that $dk_{x}/dt$ is proportional
to the electric field $E_{x}$. The second term says that an electric
field $E_{y}$ leads to a change in $k_{x}$ if there is shear in
the system. The third term is, semiclassically, the Lorentz force
- changing the polarization in the $y$ direction ($\mathcal{F}_{t\bar{y}}\neq0$)
leads to a 1D current in the $y$ direction, which then feels the
Lorentz force of the magnetic field ($\mathcal{F}_{xy}\neq0$), causing
$k_{x}$ to change.

\noindent \textbf{Charge component:}

The charge responses are
\begin{equation}
j_{2D}^{t}=\frac{1}{4\pi^{2}}\int d^{2}\bar{x}\text{tr}\left[\mathcal{F}_{xy}\mathcal{F}_{\bar{y}\bar{x}}+\mathcal{F}_{x\bar{x}}\mathcal{F}_{y\bar{y}}-\mathcal{F}_{x\bar{y}}\mathcal{F}_{y\bar{x}}\right]
\end{equation}

The first term is the Hall response for a Chern insulator. Specifically,
if $F_{xy}$ is just the magnetic field, this term becomes $j^{t}=(B_{z}/4\pi^{2})\int d^{2}\bar{x}F_{\bar{y}\bar{x}}=C_{1}B_{z}/2\pi$,
where $C_{1}$ is the first Chern number.

Consider the remaining terms
\begin{equation}
j_{2D}^{t}=\frac{1}{4\pi^{2}}\int d^{2}\bar{x}\text{tr}\left[\mathcal{F}_{x\bar{x}}\mathcal{F}_{y\bar{y}}-\mathcal{F}_{x\bar{y}}\mathcal{F}_{y\bar{x}}\right]\label{eqn:4DStrainTerms}
\end{equation}
In the simplest case of a single featureless, flat band, $f_{i\bar{j}}\propto\partial_{i}u_{j}$
where $\mathbf{u}$ is the displacement vector. This is because infinitesimal
motion $dx_{i}$ in the $i$ direction leads to a translation $dx_{j}=\partial_{i}u_{j}dx_{i}$
in the $j$ direction, which is, for Bloch wavefunctions, the same
as accumulating a $\bar{j}$-dependent phase $B_{0}\bar{j}dx_{j}$.
Hence $a_{i}=\bar{j}\partial_{i}u_{j}$ with our convention of $B_{0}=1$.

In this simple case, then, Eq. (\ref{eqn:4DStrainTerms}) becomes
\begin{equation}
j_{2D}^{t}=\frac{1}{4\pi^{2}}\int d^{2}\bar{x}\left(\left(1+\partial_{x}u_{x}\right)\left(1+\partial_{y}u_{y}\right)-\partial_{x}u_{y}\partial_{y}u_{x}\right)
\end{equation}
The expression inside the parentheses is the determinant of the deformation
gradient, that is, the area of the strained unit cell in units of
the original unit cell area. Hence the non-background terms are just
due to the change in the area of the unit cell. Adding features to
the bands will lead to corrections due to, for example, strain changing
the local density of states.

We summarize the 4D responses in Table \ref{table:4Dresponses}.

\begin{table*}
\begin{tabular}{|c|c|}
\hline
\textbf{Current density component}  & \textbf{Response}\tabularnewline
\hline
Real space  & Change in polarization\tabularnewline
 & Hall response\tabularnewline
 & Change in polarization with strain\tabularnewline
\hline
$k$-space  & Electric force\tabularnewline
 & Sheared response to electric field\tabularnewline
 & Lorentz force\tabularnewline
\hline
Charge density  & Hall response\tabularnewline
 & Change in unit cell area due to strain \tabularnewline
\hline
\end{tabular}\caption{Summary of responses from 4D phase space.}

\label{table:4Dresponses}
\end{table*}

\subsection{6D Phase Space}

\label{section:6+1Dresponses}

The action is given by
\begin{equation}
S=\frac{1}{192\pi^{3}}\int dtd^{6}x\text{tr}\left[C(x,\bar{x})\epsilon\mathcal{A}\partial\mathcal{A}\partial\mathcal{A}\partial\mathcal{A}+...\right]
\end{equation}
with $+...$ again representing the terms for a non-Abelian gauge
field. For simplicity of exposition and interpretation, we will assume
that the momentum-space and time components of $\mathcal{A}$ are
$U(N)$ and the real-space components are $U(1)$, that is, the latter
couple to all the bands in the same way. This assumption is not necessary
for our theory, however. We again assume that $C = 1$ uniformly to look at bulk responses.

There are 15 different responses in each component. If we separate
$\mathcal{F}_{x\bar{x}}$ into its background and non-background components,
for the spatial and momentum components we get an extra 7 terms for
a total of 22. For the charge component, there are 28. We sort them,
neglecting relative minus signs between the groups.

\noindent \textbf{Spatial components:}

Quasi-1D response (5 terms):
\begin{equation}
j_{3D}^{x}=\frac{1}{8\pi^{3}}\int d^{3}\bar{x}\text{tr}\left[\mathcal{F}_{t\bar{x}}\left(\left(F_{y\bar{y}}+f_{y\bar{y}}\right)\left(F_{y\bar{y}}+f_{y\bar{y}}\right)-\mathcal{F}_{y\bar{z}}\mathcal{F}_{z\bar{y}}\right)\right]
\end{equation}
By the same computation that was done for the charge response in 4D,
$\mathcal{F}_{y\bar{y}}\mathcal{F}_{z\bar{z}}-\mathcal{F}_{y\bar{z}}\mathcal{F}_{z\bar{y}}$
is the change in area perpendicular to the current. This response thus has the
form of the 1D real space response (time-varying polarization) times
the change in area perpendicular to the current.

\begin{figure}
\begin{centering}
\includegraphics[width=0.95\columnwidth]{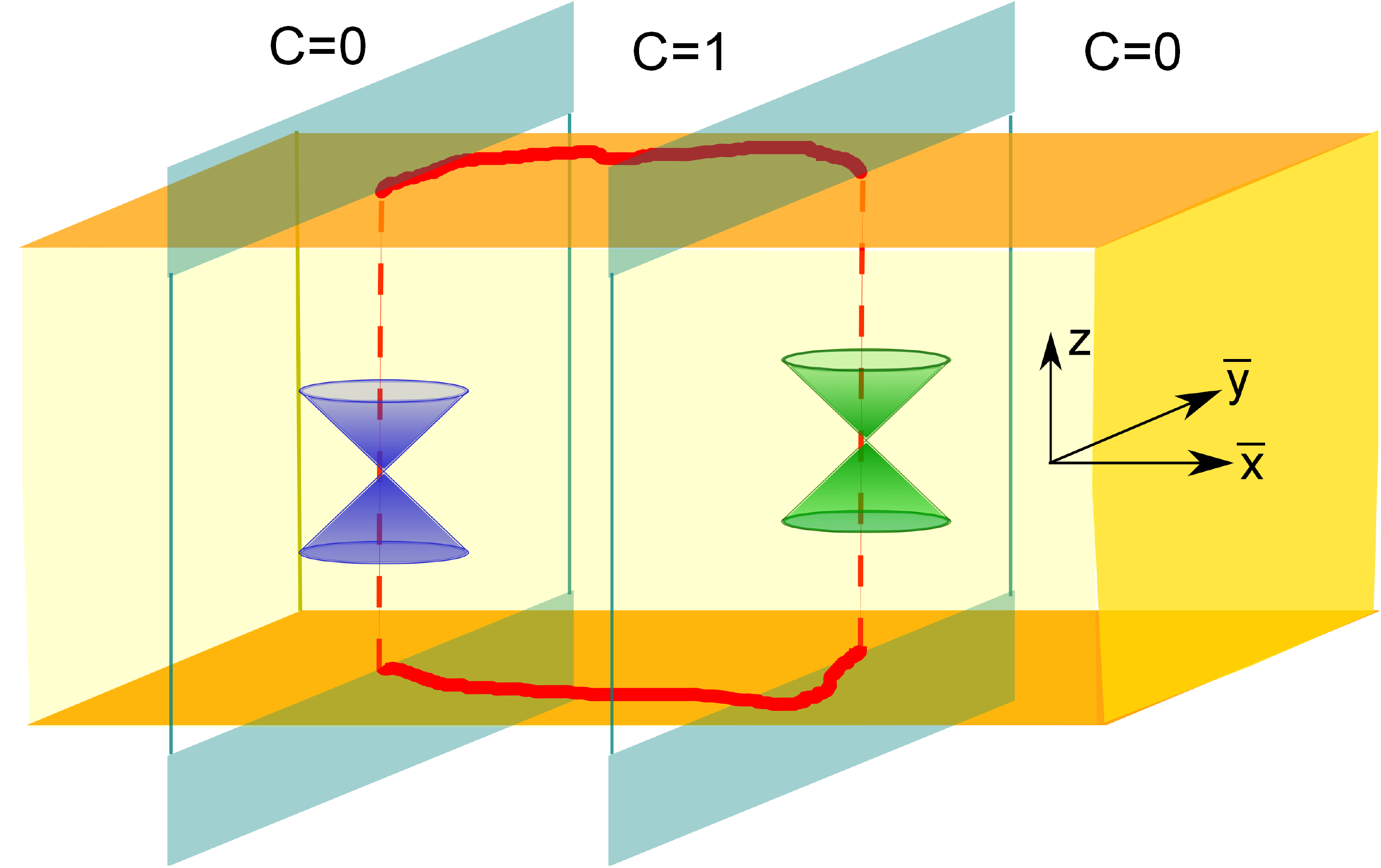}
\par\end{centering}

\caption{Slab of WSM with Weyl nodes separated along $\bar{x}$. Each slice in momentum space with fixed $\bar{x}$ can be characterized by a Chern number $C_1(\bar{x})$, which changes by unity across the Weyl nodes. Thus, the region between the nodes in the above figure is a series of Chern insulators. The edge states of these Chern insulators constitute the Fermi arcs, marked as thick red lines with an irregular shape. Note that the cones are only present as a cartoon to depict the position of the Weyl nodes; the vertical direction in the figure is $z$, and should not be confused with energy. Figure adapted from Ref. \cite{HosurWeylReview}.\label{fig:FermiArcs}}
\end{figure}

Layered Chern insulator response (2 terms):
\begin{align}
j_{3D}^{x}=\frac{1}{8\pi^{3}}\int d^{3}\bar{x}\text{tr}\left[\mathcal{F}_{ty}\mathcal{F}_{\bar{x}\bar{y}}F_{z\bar{z}}+(y\leftrightarrow z)\right]
\end{align}
where $(y\leftrightarrow z)$ means to switch $y$ and $z$ as well
as $\bar{y}$ and $\bar{z}$. This is the Hall response corresponding
to thinking of the 3D system as 2D systems layered in momentum space.
Note that this includes the Hall response of a Weyl semimetal (WSM)\cite{ZyuninBurkovWeylTheta, ChenAxionResponse,HosurWeylReview,RanQHWeyl,QiWeylAnomaly}
appearing from its monopoles of $\mathcal{F}_{\bar{x}\bar{y}}$. This
can be seen by thinking of the (2-node) WSM as stacks of 2D insulators
parametrized by the momentum direction along which the Weyl nodes
are split; as shown in Fig. \ref{fig:FermiArcs}, each insulator lying between the nodes is a Chern insulator
and thus contributes to $\mathcal{F}_{\bar{x}\bar{y}}$ (for $k_{z}$-direction
Weyl node splitting).  In this special case, integration yields
\begin{equation}
j_{3D}^{x}=\frac{1}{4\pi^{2}}\left(E_{y}\Delta k_{z}-E_{z}\Delta k_{y}\right)
\end{equation}
where $\Delta k_{i}$ is the splitting of the Weyl points in the $k_{i}$
direction.

Topological magnetoelectric effect (3 terms):
\begin{equation}
j_{3D}^{x}=\frac{1}{8\pi^{3}}\int d^{3}\bar{x}\text{tr}\left[(\mathcal{F}_{\bar{x}t}\mathcal{F}_{\bar{y}\bar{z}}-\mathcal{F}_{\bar{x}\bar{y}}\mathcal{F}_{t\bar{z}}+\mathcal{F}_{\bar{x}\bar{z}}\mathcal{F}_{t\bar{y}}\right)\mathcal{F}_{yz}]\label{eqn:unsimplifiedTME}
\end{equation}
Assuming that $\mathcal{F}_{yz}$ does not depend on momentum for
simplicity, this term is an x-direction current proportional to $B_{x}$.
Indeed, if we assume that the real space system is gapped so that
there are no monopoles of Berry curvature, simple but tedious manipulations
(see Appendix \ref{app:TME}) turn Eq. (\ref{eqn:unsimplifiedTME})
into
\begin{align}
j_{3D}^{x} & =-\frac{1}{16\pi^{3}}\partial_{t}\int d^{3}\bar{x}\epsilon^{IJK}\text{tr}\left[a_{I}\partial_{J}a_{K}+\frac{2}{3}a_{I}a_{J}a_{K}\right]B_{x}\nonumber \\
 & \equiv-\frac{1}{2\pi}(\partial_{t}P_{3})B_{x}\label{eqn:TME}
\end{align}
where $I,J,K$ run over $\bar{x},\bar{y},\bar{z}$ and $P_{3}$ is
the 3-dimensional analog of charge polarization. Eq. (\ref{eqn:TME})
is precisely the contribution of topological magnetoelectric effect
to $j^{x}$.\cite{QiHughesZhangTFT}

Topological insulator (TI)-type anomalous Hall response (6 terms):
\begin{align}
j_{3D}^{x}=\frac{1}{8\pi^{3}}\int d^{3}\bar{x}\text{tr}\big[\mathcal{F}_{tz}\left(\mathcal{F}_{\bar{x}\bar{y}}\mathcal{F}_{y\bar{z}}-\mathcal{F}_{\bar{y}\bar{z}}\mathcal{F}_{\bar{x}y}\right)\nonumber \\
+\mathcal{F}_{ty}\mathcal{F}_{\bar{x}\bar{y}}f_{z\bar{z}}+(y\leftrightarrow z)\big]\label{eqn:strainedChernResp}
\end{align}

If we choose a gauge such that the real space Berry connections do
not depend on momentum, then by similar logic to the topological magnetoelectric
effect terms (and with similar assumptions), we can manipulate this
contribution into the form
\begin{equation}
j_{3D}^{x}=-\frac{1}{2\pi}\left(E_{y}\partial_{y}-E_{z}\partial_{z}\right)P_{3}
\end{equation}

Here $E$ is the real space electric field. This is the anomalous
Hall effect that appears in a 3D TI.\cite{QiHughesZhangTFT} It differs
from the Hall effect that appears as a quasi-2D response in that it
does not originate from having a nonzero total Chern number at each
2D slice of momentum space.

Sheared polarization responses (6 terms):
\begin{align}
j_{3D}^{x} & =\frac{1}{8\pi^{3}}\int & d^{3}\bar{x}\text{tr}\left[\mathcal{F}_{t\bar{y}}\left(\mathcal{F}_{\bar{x}z}\mathcal{F}_{\bar{z}y}+\mathcal{F}_{\bar{x}y}\left(F_{z\bar{z}}+f_{z\bar{z}}\right)\right)\right.\nonumber \\
 & \left.+(y\leftrightarrow z)\right] 
\end{align}
The first term here corresponds to a current flowing in $y$ due to
a changing polarization, but that current being redirected into the
$z$ and then the $x$ direction by shear. The second term is the
same current in $y$ being redirected into the $x$ direction together
with a change in the perpendicular area due to uniaxial strain. These are the 3D real space analogues of the 2D real space response illustrated in Fig. \ref{fig:strainedPolarization}.

\noindent $\mathbf{k}$\textbf{-space components:}

Quasi-1D response (5 terms):
\begin{align}
j^{\bar{x}}=\frac{1}{8\pi^{3}}\text{tr}\left[\mathcal{F}_{tx}\left((F_{y\bar{y}}+f_{y\bar{y}})(F_{z\bar{z}}+f_{z\bar{z}})-\mathcal{F}_{y\bar{z}}\mathcal{F}_{z\bar{y}}\right)\right]
\end{align}
As in the real space response, this is the 1D response accounting
for changes in the area perpendicular to the current.

Strained electric forces (6 terms):
\begin{align}
j^{\bar{x}}=\frac{1}{8\pi^{3}}\text{tr} & \left[\mathcal{F}_{ty}\left(\mathcal{F}_{x\bar{z}}\mathcal{F}_{\bar{y}z}+\mathcal{F}_{x\bar{y}}(F_{z\bar{z}}+f_{z\bar{z}})\right)+(y\leftrightarrow z)\right]
\end{align}

These terms correspond to a typical electrical force in the $y$ direction,
which is then redirected to the $x$ direction by shears and correcting
for change in the area perpendicular to the current.

Strained polarization/Lorentz force responses (8 terms):
\begin{align}
j^{\bar{x}}=\frac{1}{8\pi^{3}}\text{tr} & \left[\mathcal{F}_{t\bar{y}}\left(\mathcal{F}_{zx}\mathcal{F}_{y\bar{z}}+\mathcal{F}_{x\bar{z}}\mathcal{F}_{yz}+\mathcal{F}_{xy}(F_{z\bar{z}}+f_{z\bar{z}})\right)\right.\nonumber \\
 & \left.+(y\leftrightarrow z)\right]
\end{align}

The first two terms say that if the polarization changes in the $y$
direction, then either shear or the Lorentz force can change this
into a current in the $z$ direction. That current can then be redirected
by the Lorentz force or shear (respectively) to the $x$ direction.
The third term is the same current due to polarization, leading to
a current in the $x$ direction by the Lorentz force, correcting for
change in the area perpendicular to the new current.

WSM-type $\mathbf{E}\cdot\mathbf{B}$ charge pumping (3 terms):
\begin{equation}
j^{\bar{x}}=\frac{1}{8\pi^{3}}\text{tr}\left[\left(\mathcal{F}_{tx}\mathcal{F}_{yz}+\mathcal{F}_{ty}\mathcal{F}_{zx}+\mathcal{F}_{tz}\mathcal{F}_{xy}\right)\mathcal{F}_{\bar{y}\bar{z}}\right]
\end{equation}

Assuming the real-space field strengths are $k$-independent, then
this term is $\mathbf{E}\cdot\mathbf{B}$ times the Berry curvature.
If we integrate over $\bar{y}$ and $\bar{z}$, we find
\begin{equation}
\int d\bar{y}d\bar{z}j^{\bar{x}}=\frac{1}{4\pi^{2}}C_{1}(\bar{x})\mathbf{E}\cdot\mathbf{B}
\end{equation}
where $C_{1}(\bar{x})$ is the Chern number of the slice of the Brillouin
zone at fixed $\bar{x}$. For the case of a WSM with Weyl points split
along the $\bar{x}$ direction, $C_{1}(\bar{x})$ is nonzero between
the Weyl points (see Fig. \ref{fig:FermiArcs}), so this response is a current from one Weyl point
to the other . This is exactly the chiral anomaly\cite{ZyuninBurkovWeylTheta,ChenAxionResponse, HosurWeylReview,RanQHWeyl,QiWeylAnomaly,NielsenABJ,VolovikBook}, which says that
an $\mathbf{E}\cdot\mathbf{B}$ field in a WSM pumps charge from one
Weyl point to the other.

\noindent \textbf{Charge component:}

All 16 terms which only contain mixed field strengths $\mathcal{F}_{i\bar{j}}$
combine to form the unit cell volume, corrected for strain. The other
two types of response are:

Layered Chern insulator Hall response (3 terms):
\begin{equation}
j^{0}=\frac{1}{8\pi^{3}}\int d^{3}\bar{x}\text{tr}\left[\mathcal{F}_{xz}\mathcal{F}_{\bar{z}\bar{x}}F_{y\bar{y}}+\text{ perm.}\right]
\end{equation}
where ``perm.'' indicates terms created by cyclically permuting
$(x,y,z)$ and $(\bar{x},\bar{y},\bar{z})$. This is the charge density
counterpart to the spatial layered Chern insulator Hall response;
it analogously comes from adding the Hall response of each subsystem
at fixed $k_{i}$ to a magnetic field $B_{i}$.

TI-type Hall response (9 terms):
\begin{align}
j^{0}=\frac{1}{8\pi^{3}}\int d^{3}\bar{x}\text{tr} & \left[\mathcal{F}_{xz}\left(\mathcal{F}_{\bar{y}\bar{z}}\mathcal{F}_{y\bar{x}}+\mathcal{F}_{\bar{x}\bar{y}}\mathcal{F}_{y\bar{z}}+\mathcal{F}_{\bar{z}\bar{x}}f_{y\bar{y}}\right)\right.\nonumber \\
 & \left.+\text{ perm.}\right]
\end{align}
By the same methods used to derive Eq. (\ref{eqn:TME}), these terms
can be massaged into the form
\begin{equation}
j^{0}=\frac{1}{2\pi}\mathbf{B}\cdot\nabla P_{3}
\end{equation}
This is exactly the charge component of the Hall response that appears
in a TI\cite{QiHughesZhangTFT}.

We summarize the 6D responses in Table \ref{table:6Dresponses}.

\begin{table*}
\begin{tabular}{|c|c|}
\hline
\textbf{Current component}  & \textbf{Response}\tabularnewline
\hline
Real space  & Quasi-1D responses\tabularnewline
 & Quasi-2D (layered Chern insulator) response\tabularnewline
 & Topological magnetoelectric effect\tabularnewline
 & TI-like anomalous Hall response\tabularnewline
 & Change in polarization with strain\tabularnewline
\hline
$k$-space  & Quasi-1D response\tabularnewline
 & Electric fields with strain\tabularnewline
 & Change in polarization plus Lorentz force with strain\tabularnewline
 & WSM-like $\mathbf{E}\cdot\mathbf{B}$ charge pumping\tabularnewline
\hline
Charge density  & Density response to change in unit cell volume \tabularnewline
 & Layered Chern insulator Hall response \tabularnewline
 & TI-like anomalous Hall response\tabularnewline
\hline
\end{tabular}\caption{Summary of 6D phase space responses.}

\label{table:6Dresponses}
\end{table*}

\section{Anomalies}

\label{section:anomalies}

In the previous section, we enumerated the phase space bulk responses,
which, as we have seen, correspond to the topological responses of
filled states in real space. This includes the responses of insulators, semimetals and the responses of metals that involve all the occupied states, such as the anomalous Hall effect. We now wish to describe the
topological features of Fermi surfaces and real space system edges. We will
also approach the response of semimetals from another perspective.
All these features take the form of anomalies in phase space.

Given the CS theory for a phase space system, such as that which appears
in Eq. (\ref{eqn:ChernSimons2D}) or its higher-dimensional and/or
non-Abelian generalization, suppose that $\partial_{i}C\neq0$ for
some coordinate $i$. This means that the phase space system contains
an edge, that is, the real space system has a Fermi surface or an
edge. Then in general, the responses of the system will depend on
details; for example, edge currents of quantum Hall systems depend
on the non-universal edge mode velocity. However, there will be a
universal anomaly (or lack thereof) along such edges. We see this
from the anomaly resulting from the phase-space CS term in $2n$D:
\begin{equation}
\sum_{\mu\neq i}\partial_{\mu}j^{\mu}=\frac{1}{n!2^{2n}\pi^{n}}\partial_{i}C\epsilon^{i\alpha_{1}\alpha_{2}...\alpha_{2n}}\text{tr}\left[\mathcal{F}_{\alpha_{1}\alpha_{2}}...\mathcal{F}_{\alpha_{2n-1}\alpha_{2n}}\right]\label{eq:anomaly-general}
\end{equation}
From here, integration over the appropriate phase space directions
determines the anomalies in the real system. We list some physically
interesting effects below.

\subsection{Fermi Surface Anomalies}

In Section \ref{section:2DExample}, we computed the chiral anomaly
in 1D momentum space by imposing edges at $\bar{x}=\pm k_{F}$ in
2D phase space. Integrating over $x$ yielded the fact that an electric
field pumps electrons in states near $+k_{F}$ into states near $-k_{F}$,
or vice versa. At a down-to-earth level, this simply corresponds to
tilting of the 1D Fermi surface in an electric field due to semiclassical
motion of the electrons.

This idea is straightforward to generalize to higher dimensions. For
instance, an open Fermi surface in 2D is obtained by confining the
4D phase space system between $\bar{x}=\pm k_{F}$ while leaving the
other three directions infinite, while a 3D spherical Fermi surface
results from making the $\bar{x}$, $\bar{y}$ and $\bar{z}$ directions
in 6D phase space finite under the constraint $\bar{x}^{2}+\bar{y}^{2}+\bar{z}^{2}=k_{F}^{2}$
and leaving the $x$, $y$ and $z$ directions unconstrained. For
each phase space geometry, the corresponding anomaly characterizes
properties of the resultant Fermi surface. Importantly, if a special
object such as a Dirac or a Weyl point is buried under the Fermi surface,
its observable effects in local transport phenomenon should emerge
from the anomaly equation.

We demonstrate this first for a 3D spherical Fermi surface, which
carries a Chern number in general and exhibits a chiral anomaly proportional
to the Chern number and the electromagnetic field $\mathbf{E}\cdot \mathbf{B}$.
The most well-known occurence of this phenomenon is in Weyl semimetals.
We begin with the anomaly equation in 6D phase space. In the absence
of any strains and ignoring quasi-lower-dimensional terms (i.e., terms
such as $F_{x\bar{x}}$ that contain the background field), it reads
\begin{alignat}{1}
 & \sum_{\mu}\partial_{\mu}j^{\mu}-\partial_{\bar{r}}j^{\bar{r}}=\nonumber \\
 & \,\,\,\frac{1}{8\pi^{3}}\delta(\bar{r}-k_{F})\mathcal{F}_{\bar{\theta}\bar{\phi}}\left(\mathcal{F}_{tx}\mathcal{F}_{yz}+\mathcal{F}_{ty}\mathcal{F}_{zx}+\mathcal{F}_{tz}\mathcal{F}_{xy}\right)
\end{alignat}
where $(\bar{r},\bar{\theta},\bar{\phi})$ are the spherical
co-ordinates corresponding to $(\bar{x},\overline{y},\overline{z})$.
Integrating over the barred co-ordinates immediately yields the chiral
anomaly in Weyl semimetals:
\begin{equation}
\sum_{\mu=t,x,y,z}\partial_{\mu}j_{3D}^{\mu}=\frac{1}{4\pi^2}C_{FS}\mathbf{E}\cdot\mathbf{B}\label{eq:Weyl-anomaly}
\end{equation}
where $C_{FS}=\frac{1}{2\pi}\oint\mathrm{d}\overline{\theta}\mathrm{d}\overline{\phi}\sin\overline{\theta}\mathcal{F}_{\overline{\theta}\overline{\phi}}\in\mathbb{Z}$
is the Chern number of the Fermi surface and equals the total chirality
of all Weyl points enclosed by it.

Unlike in 3D, in 2D systems the one-dimensional Fermi surface carries a non-quantized Berry phase
instead of a Chern number. An analogous analysis, i.e.,
starting with the anomaly equation in 4D phase space with a $(\overline{x},\overline{y})$
boundary that satisfies $\overline{x}^{2}+\overline{y}^{2}=k_{F}^{2}$
and integrating over $(\overline{x},\overline{y})$ gives
\begin{equation}
\sum_{\mu=t,x,y}\partial_{\mu}j_{2D}^{\mu}=\frac{1}{4\pi^2}B_{z}\partial_{t}\gamma\label{eq:AH-metal}
\end{equation}
ignoring strain and quasi-lower-dimensional terms, where $\gamma=\ointop_{FS}\mathbf{a_{\bar{r}}}\cdot\mathrm{d}\mathbf{\bar{r}}$
is the Berry phase on the Fermi surface. Eq (\ref{eq:AH-metal}) is
the statement that adiabatically changing the Hall conductivity of
an anomalous Hall metal in a magnetic field creates charged excitations
bound to the field.

In the presence of strains, both (\ref{eq:Weyl-anomaly}) and (\ref{eq:AH-metal})
contain more terms on their right hand sides. We will encounter these
terms in the next subsection when we discuss the effects of dislocations.
Before moving on, however, we wish to stress that the physical anomaly
in a given dimension is independent of the topology of the
Fermi surface. However, certain topologies are more convenient for
studying a given physical anomaly. For instance, the chiral anomaly
in Weyl metals is easier to see for a spherical Fermi surface, but
it can be equally well be derived for open Fermi surfaces that span
one or two directions in the Brillouin zone.

\subsection{Anomalies in Real Space}

\subsubsection{Real space edge}

The simplest example of a real space edge anomaly was derived in Sec
\ref{section:2DExample}, where we imposed $x$-direction edges in
2D phase space and showed that a time-dependent polarization in 1D
can be used to pump charge across the length of the chain. As a more
non-trival example, consider a real, single-band 2D Chern insulator
which occupies $x>0$. Then in 4D phase space, the anomaly equation
(\ref{eq:anomaly-general}) with $C(x)=\Theta(x)$ reads
\begin{align}
\partial_{t}\rho & +\partial_{y}j^{y}+\partial_{\bar{x}}j^{\bar{x}}+\partial_{\bar{y}}j^{\bar{y}}=\nonumber \\
 & \frac{1}{4\pi^{2}}\delta(x)\left(\mathcal{F}_{ty}\mathcal{F}_{\bar{x}\bar{y}}-\mathcal{F}_{t\bar{x}}\mathcal{F}_{y\bar{y}}+\mathcal{F}_{t\bar{y}}\mathcal{F}_{y\bar{x}}\right)
\end{align}
Integrating $\partial_{\bar{x}}j^{\bar{x}}+\partial_{\bar{y}}j^{\bar{y}}$
over momentum space gives zero since there is no boundary in those
directions. Hence, integrating the previous equation over momentum
space gives
\begin{equation}
\partial_{t}\rho_{2D}+\partial_{y}j_{2D}^{y}=\frac{1}{4\pi^{2}}\delta(x)\int d^{2}k\mathcal{F}_{\bar{x}\bar{y}}E_{y}=\frac{1}{2\pi}\delta(x)C_{1}E_{y}\label{eqn:ChernAnomaly}
\end{equation}
with $C_{1}$ the first Chern number of the occupied band of the 2D
Hamiltonian. We have ignored quasi-1D terms and terms containing strain.
Eq. (\ref{eqn:ChernAnomaly}) is recognizable as the usual anomaly
for a 2D Chern insulator where an electric field parallel to the edge
builds up a charge density along that edge.

\subsubsection{Dislocations}

\begin{figure}
\begin{centering}
\includegraphics[width=0.8\columnwidth]{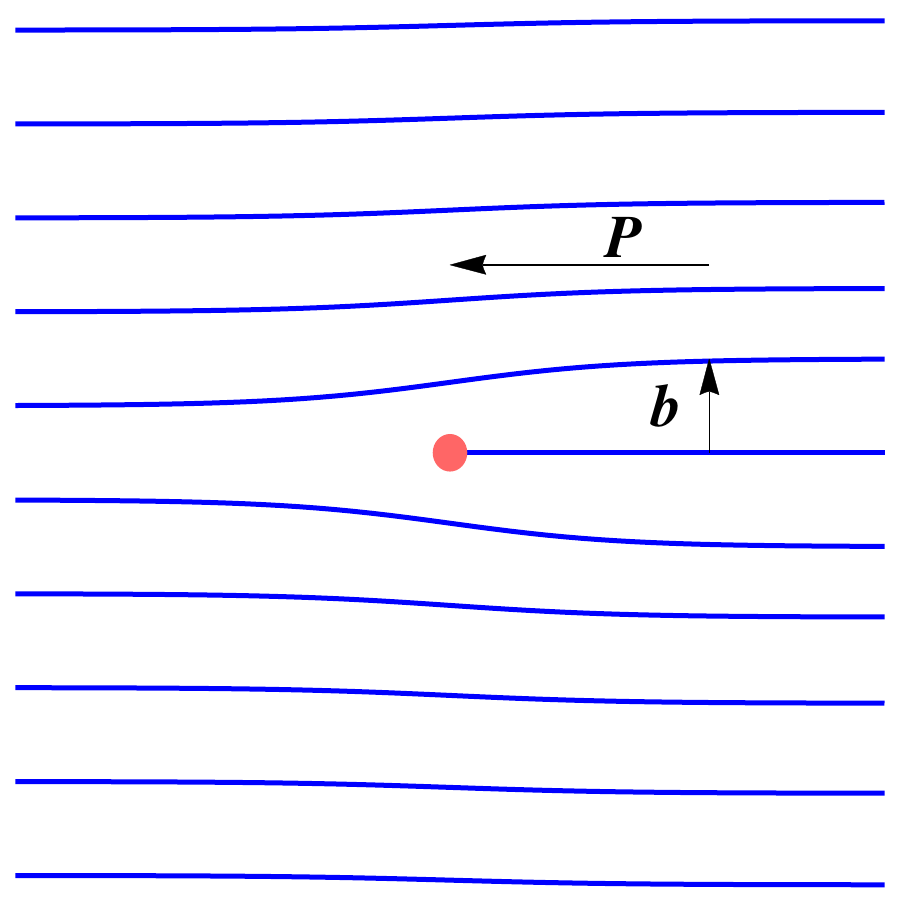}
\par\end{centering}

\caption{An edge dislocation in 2D with Burger's vector $\mathbf{b}$.
In the presence of a polarization $\mathbf{P}\perp\mathbf{b}$,
charge gets accumulated at the core of the dislocation, shown by the
red dot.\label{fig:2D-edge-disloc}}
\end{figure}

\textbf{4D phase space:} The simplest example of a dislocation is
an edge dislocation in 2D real space. The key feature of the dislocation,
as we discussed in Section \ref{subsection:interpretation}, is that,
far from the dislocation line itself, electrons accumulate a Berry's
phase of $\mathbf{k}\cdot\mathbf{b}$ upon encircling the
dislocation. We can thus model the dislocation by a Berry connection
$(a_{r},a_{\theta})=(0,\frac{\mathbf{b}\cdot\mathbf{k}}{2\pi})$,
leading to a $\mathbf{k}$-independent Berry curvature $\mathcal{F}_{\theta\bar{i}}=b_{i}/2\pi r$.
Our theory breaks down at the dislocation itself because the system
changes quickly on the scale of a lattice constant. We can avoid this
problem by keeping the Berry connection but surrounding the dislocation
by a finite size puncture in the system of radius $r_{0}$, i.e. choose
$C=\Theta(r-r_{0})$ with $r$ the radial coordinate in the $xy$-plane.
The resulting anomaly equation reads
\begin{equation}
\sum_{\mu}\partial_{\mu}j^{\mu}-\partial_{r}j^{r}=\frac{1}{8\pi^3}\left(\mathcal{F}_{t\overline{x}}b_{y}-\mathcal{F}_{t\overline{y}}b_{x}\right)
\end{equation}
plus quasi-1D terms on the right-hand-side, which we ignore. Integrating
over $\overline{x},\overline{y}$ and $\theta$ gives the charge radiating
from the core of an edge dislocation in the presence of a time-dependent
polarization:
\begin{equation}
\partial_{t}\rho_{2D}=\hat{\mathbf{z}}\cdot\left(\mathbf{b}\times\partial_{t}\mathbf{P}\right)
\end{equation}
This is shown in Fig \ref{fig:2D-edge-disloc}, which makes the physical
picture of the anomaly clear in the limit of weakly coupled chains
perpendicular to $\mathbf{b}$; the core of the dislocation is
the end of such a chain, so polarizing that chain adds charge to its end.
The non-trivial result is that the extra charge remains bound to the
dislocation core and does not leak into other chains even when they
are strongly coupled.

\textbf{6D phase space: }A similar analysis for a disocation in 3D
real space running along $\hat{\mathbf{z}}$ and with Burger's vector
$\mathbf{b}$ gives
\begin{align}
 & \partial_{t}\rho+\frac{1}{r}\partial_{\theta}j^{\theta}+\partial_{z}j^{z}\nonumber \\
 & =\frac{\delta(r-r_{0})}{8\pi^{3}}\int d^{3}\bar{x}\left(\mathcal{F}_{\bar{x}\bar{y}}(r\mathcal{F}_{\theta\bar{z}})+\mathcal{F}_{\bar{y}\bar{z}}(r\mathcal{F}_{\theta\bar{x}})+\mathcal{F}_{\bar{z}\bar{x}}(r\mathcal{F}_{\theta\bar{y}})\right)\mathcal{F}_{tz}\nonumber \\
 & =-\frac{\delta(r-r_{0})}{8\pi^{3}}E_{z}\int d^{3}\bar{x}\mathbf{\Omega}\cdot\mathbf{b}\label{eq:3D-disloc-anomaly}
\end{align}
where $\Omega_{i}=\frac{1}{2}\epsilon_{ijk}\mathcal{F}_{\bar{j}\bar{k}}$
is the Berry curvature of the bands in the plane perpendicular to
$i$.

\begin{figure}
\centering{}\includegraphics[clip,width=0.5\columnwidth,height=4.5cm]{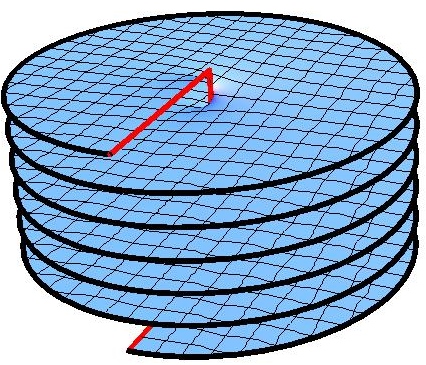}\includegraphics[width=0.5\columnwidth]{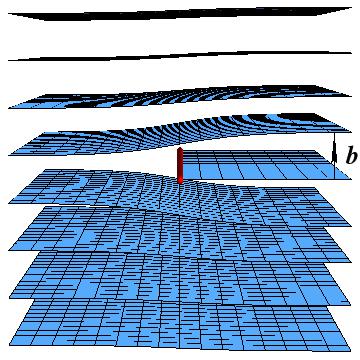}\caption{Screw (left) and edge (right) dislocations in 3D. Dislocations in
an axion insulator harbor chiral modes, denoted by red lines in both
figures. In the screw dislocation, thick black lines represent the
standard chiral edge mode. The screw dislocation geometry with Weyl
nodes split along the screw axis was used for the numerical results
presented in Fig \ref{fig:Disloc-numerics}.\label{fig:3D-disloc}}
\end{figure}

To understand (\ref{eq:3D-disloc-anomaly}), let us first consider
a layered Chern insulator, that is, a system composed of layers of
Chern insulators stacked along a certain direction. The integral in
(\ref{eq:3D-disloc-anomaly}) then gives the Chern number of the layers
in each direction, so
\begin{equation}
\partial_{t}\rho+\frac{1}{r}\partial_{\theta}j^{\theta}+\partial_{z}j^{z}=-\frac{\delta(r-r_{0})}{\pi}E_{z}\mathbf{C}\cdot\mathbf{b}\label{eq:Chern-layers-disloc}
\end{equation}
where $C_{i}=\frac{1}{8\pi^{2}}\int\mathrm{d}^{3}\overline{x}\Omega_{i}$.
Now, add a dislocation running along $\hat{\mathbf{z}}$ with Burger's
vector $\mathbf{b}$. Edge dislocations are defined by $\mathbf{b}\perp\hat{\mathbf{z}}$
whereas screw dislocations have $\mathbf{b}\parallel\hat{\mathbf{z}}$.
The two scenarios are shown in Fig \ref{fig:3D-disloc}. Eq (\ref{eq:Chern-layers-disloc})
says that in either case, there exists a chiral mode along the dislocation
that participates in an anomaly in response to $E_{z}$. We can understand
this as follows.

An edge dislocation can be thought of as a semi-infinite sheet perpendicular
to $\mathbf{b}$ and unbounded along $\hat{\mathbf{z}}$ inserted
into the 3D lattice. If the sheet has a Chern number, we expect it
to have a chiral mode along $\hat{\mathbf{z}}$. For weakly coupled
sheets, this is precisely the chiral mode along the edge dislocation.
For a screw dislocation, the existence of a chiral dislocation mode
follows from an argument adapted from one that predicts helical dislocation
modes in weak topological insulators\cite{VishwanathWeakDisloc, HughesMajoranaDisloc}.
Suppose that our system is of finite size in the $z$ direction. Then
on each surface, there is a semi-infinite edge emerging from the dislocation.
However, this edge must carry a chiral mode since the surface layer
is a Chern insulator. By charge conservation, this chiral mode cannot
terminate at the dislocation, so the chiral mode must proceed along
the dislocation to the other surface. Moreover, in each case, the
chiral mode is expected to survive for strongly coupled layers as
well, where the system is better thought of as stacked sheets in momentum
space and is typically termed an axion insulator. This is because
layered Chern insulators and axion insulators are actually the same
phase -- there is no phase transition as the interlayer coupling is
strengthened -- so their topological defects such as dislocations
have qualitatively similar behavior. Indeed, the presence of a chiral
mode was shown explicitly for an axion insulator created from a charge
density wave instability of a WSM in Ref. \onlinecite{Wang2013}.

\textbf{Spectral flow due to dislocations in Weyl semimetals:} Having
seen examples of anomalies being the universal feature of gapless
systems, we use our theory's anomaly machinery to derive a new result:
prediction of an anomaly at dislocation lines in a WSM. This is closely
related to the case of a layered Chern insulator (axion insulator)
just discussed.

Consider a WSM with two nodes split by $\mathbf{K}$ (and thus
having broken time reversal symmetry) with a dislocation along $z$.
In contrast to the layered Chern insulators or axion insulators, WSMs
have a gapless bulk and thus, cannot support localized modes the same
way that the former do. However, our theory allows us to confirm that there is indeed an
anomaly at the dislocation in a WSM. The anomaly calculation is identical
to the axion insulator case, except that $\frac{1}{4\pi}\int\mathrm{d}^{3}x\Omega_{i}=K_{i}$
instead of is $2\pi$. The result is that the anomaly is $\delta(r-r_{0})E_{z}\mathbf{K}\cdot\mathbf{b}/2\pi$,
which reflects the fact that chiral modes appear only in the region
of momentum space between the Weyl nodes, where the Chern number of
the layers is $\pm1$. From now on, we assume $\mathbf{K}=K\hat{\mathbf{z}}$
for concreteness.

The physical interpretation of this anomaly is more subtle for the
WSM than the axion insulator. In the latter case, due to the bulk
gap, the anomaly means that there is a chiral zero mode on the dislocation.
In the WSM case, there is no bulk gap. Furthermore, if the region
carrying a nonzero Chern number is near $k_{z}=0$, then that region
sees only small perturbations from the dislocation because the dislocation
acts like a flux proportional to $k_{z}$. Hence we should not necessarily
expect a zero mode on the dislocation. On the other hand, if this region is located near $k_{z}=\pi$, then there may be such a zero
mode. In general, however, the existence of a localized zero mode
is not guaranteed.

Since the anomaly need not imply a localized zero mode, we numerically
solved a simple $\mathbf{k}\cdot\mathbf{p}$ model for a WSM in the
presence of a dislocation in order to directly verify the anomaly.
The Hamiltonian we used is
\begin{align}
H= & \left[M_{0}+M_{1}k_{z}^{2}+M_{2}(k_{x}^{2}+k_{y}^{2})\right]\Gamma_{5}+L_{1}k_{z}\Gamma_{4}\nonumber \\
 & +L_{2}\left(k_{y}\Gamma_{1}-k_{x}\Gamma_{2}\right)+U_{0}\Gamma_{12}\label{eqn:WSMHam}
\end{align}
Here the anticommuting $\Gamma$ matrices are defined by $\Gamma_{1,2,3}=\sigma_{x,y,z}\tau_{x}$,
$\Gamma_{4}=\tau_{y}$, $\Gamma_{5}=\tau_{z}$, and $\Gamma_{ij}=[\Gamma_{i},\Gamma_{j}]/2i$
where $\sigma$ is a spin index and $\tau$ is an orbital index. This
model leads to Weyl points at $\mathbf{k}=\pm|U_{0}|/L_{1}\mathbf{\hat{z}}$
when the quadratic term is neglected. This model has been previously
investigated in a radial geometry\cite{QiWeylAnomaly} with no dislocation.
The only effect of a screw dislocation at $r=0$ with Burgers vector
$b\mathbf{\hat{z}}$ is that the dependence of the components of the
wavefunction on the in-plane angle $\theta$ changes from $e^{in\theta}$
to $e^{i(l+bk_{z}/2\pi)\theta}$, where the half-integer $l$ is the
eigenvalue of $L_{z}$ in the absence of the dislocation.

\begin{figure}
\subfigure[ ]{ \includegraphics[width=4cm]{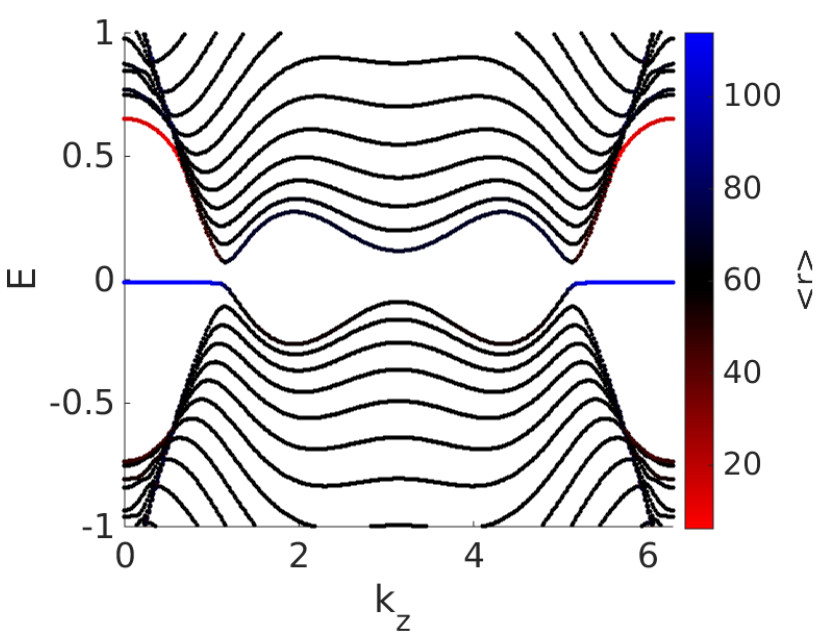}
\label{fig:WSMNoDisloc} } \subfigure[ ]{ \includegraphics[width=4cm]{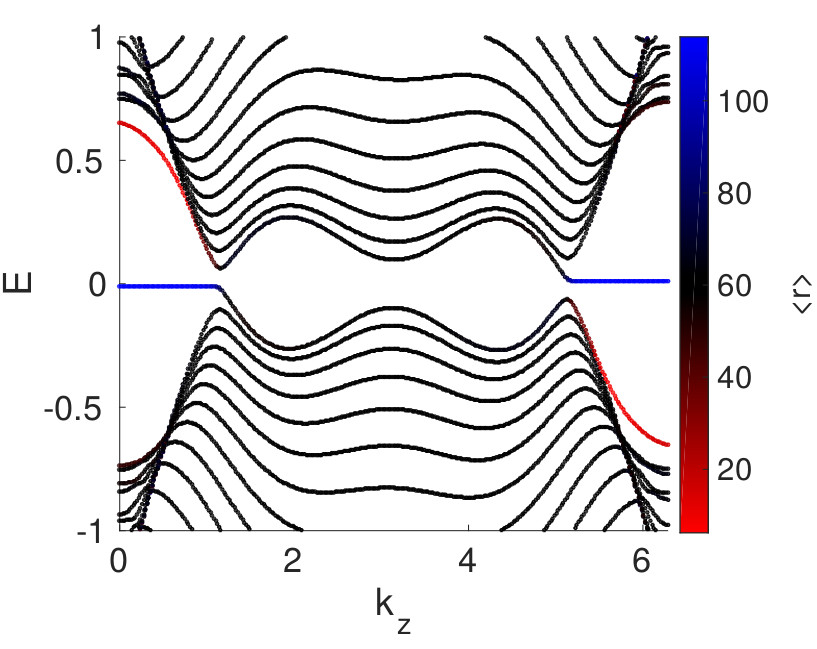}
\label{fig:WSMDisloc}} \subfigure[ ]{ \includegraphics[width=4cm]{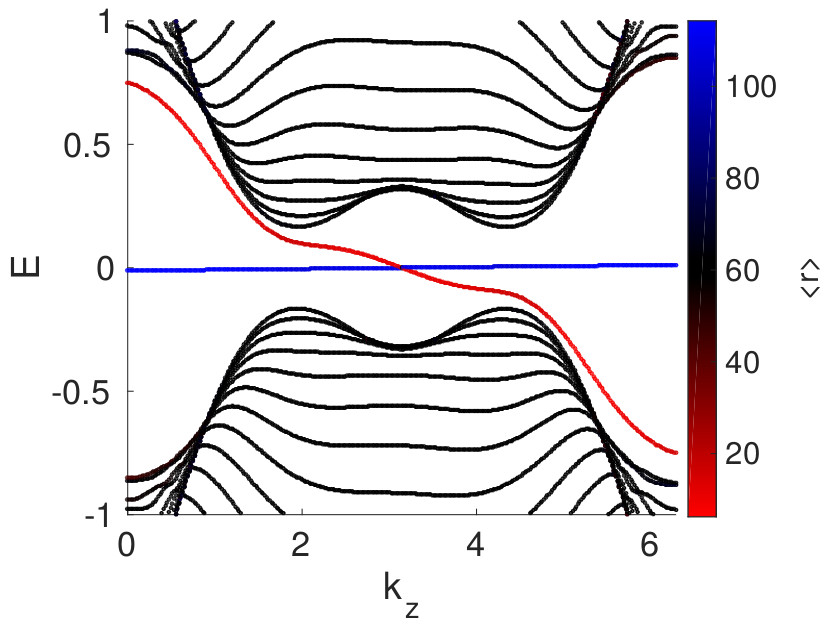}
\label{fig:axionInsulatorDisloc} } \subfigure[ ]{ \includegraphics[width=4cm]{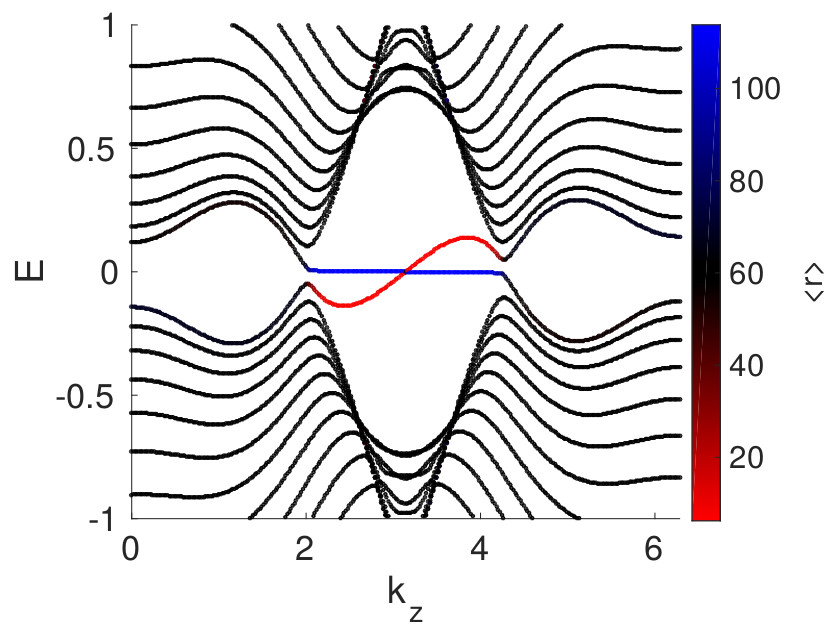}
\label{fig:WSMDislocNearPi}} \caption{Band structure of the lattice regularized version of Eqn. (\ref{eqn:WSMHam})
in a cylindrical geometry. Color corresponds to $\langle r\rangle$
with the dislocation at $r=0$; red indicates localization on the
dislocation and blue is localization on the outer boundary. Parameters
are $M_{0}=0$, $M_{1}=0.342$ eV \AA{}$^{2}$, $M_{2}=18.25$ eV
\AA{}$^{2}$, $L_{1}=1.33$ eV \AA{}, $L_{2}=2.82$ eV \AA{}, $R=120$
radial sites, $l=1/2$ angular momentum unless otherwise stated. Note
that due to the dislocation the system is not periodic in $k_{z}$
at fixed angular momentum. (a) WSM phase ($U_{0}=1.3$ eV), no dislocation.
(b) Same as (a), but with dislocation. (c) Axion insulator phase ($U_{0}=1.7$
eV) with dislocation. (d) WSM phase ($U_{0}=-1.3$ eV, $M_{1}=-0.342$
eV \AA{}, $M_{0}=1.4$ eV) with topologically nontrivial
BZ slices centered at $k_{z}=\pi$ and a dislocation. \label{fig:Disloc-numerics}}
\end{figure}

We solved the discretized version of this model at fixed angular momentum
for a cylinder of size $R=120$ sites at fixed angular momentum $l=1/2+k_{z}/2\pi$.
For comparison, we show the band structure in the WSM phase with no
dislocation in Fig. \ref{fig:WSMNoDisloc}. The mode localized near
$r=0$ (in blue) is always at higher energy than the Fermi arc, and
is not topological. Adding the dislocation, we see in Fig. \ref{fig:WSMDisloc}
that now the $r=0$ mode changes from unoccupied to occupied after
crossing the Weyl points; an electron has been pumped from the Fermi
arc (in red) to the dislocation. This is the anomaly that we discussed
above, even though there is no zero energy mode localized on the dislocation.
This system can smoothly evolve, by bringing the Weyl points together
and annihilating them, into the axion insulator in Fig. \ref{fig:axionInsulatorDisloc}.
That state has a single chiral mode localized on the dislocation which
crosses the bandgap without mixing with the outer edge mode, as expected.
In Fig. \ref{fig:WSMDislocNearPi}, we have a WSM with a topologically
nontrivial region centered about $k_{z}=\pi$; here there is a zero
mode localized on the dislocation, and the charge pumping is more
obvious than in Fig. \ref{fig:WSMDisloc}.

The result of charge pumping due to disclinations has been previously predicted\cite{jianWSMdisloc}. However, our picture is different from than the one considered there. The claim in Ref. \cite{jianWSMdisloc} is that a chiral magnetic field, which in our case is created by the dislocation, causes a net spontaneous current to flow. As can be seen from our picture, this is not true; an electric field is necessary to have an anomaly and thus a net current. Fundamentally, the total current must vanish in the absence of an electric field. If the current did not vanish, adding an electric field parallel to the current would cause dissipation and lower the system energy, but this is impossible for a system already in its ground state. The difference in Ref. \cite{jianWSMdisloc} stems from neglecting momentum space regions away from the Weyl nodes and the real space boundary in determining the total current. Thus, while the general expression for the current density derived by Ref. \cite{jianWSMdisloc} is correct, the total current vanishes when these contributions are included. For the case which we show in Fig. \ref{fig:WSMDisloc}, the net current due to the dislocation is cancelled by the current along the Fermi arcs. In the case of Fig. \ref{fig:WSMDislocNearPi}, the dislocation mode near one Weyl point connects directly to the mode on the other side through a zero mode which cancels the net current. 

To summarize, dislocations in a WSM indeed cause pumping of charge
to (or from) the dislocation line when an electric field is applied
along the dislocation. Such a charge pumping is smoothly connected
to that which occurs in the axion insulator, but it may or may not,
depending on details, result in a zero mode localized on the dislocation.
Although we presented numerics for a screw dislocation that runs along
the same direction as the Weyl node splitting, (\ref{eq:3D-disloc-anomaly})
and hence the qualitative result is valid for edge dislocations as
well as for other directions of the Weyl node splitting.

\section{Discussion and Conclusions}

\label{section:discussion}

We have shown that the responses and anomalies of a gapped or gapless
system living in $n$ spatial dimensions can be described by a single
response theory of a gapped system living in $2n$ spatial dimensions.
Conceptually, this is because adding magnetic fields in the $2n$ dimensional system and projecting onto the zeroth Landau level allows us
to interpret that system as living in phase space. We have used this
theory to reproduce well-understood responses and anomalies in systems
with non-interacting electrons and Abelian real space gauge fields,
as well as to demonstrate the existence of spectral flow due to dislocations
in Weyl semimetals.

There are several interesting fundamental questions about our theory which are at present open. It would be interesting to see how our theory connects to the use of phase space in statistical mechanics; how might the Landau-Boltzmann transport equation, that describes transport in Fermi liquids via Wigner functions, or Liouville's theorem, that describes the time-evolution of general classical systems in phase space via a density matrix, arise in our context? Both the Wigner function and the phase space density matrix treat real and momentum space on an equal footing; thus, our theory holds promise in capturing these phenomena.

In addition to these fundamental questions, we envision a number of extensions of our theory to more complicated systems. In particular, the responses that we have explicitly discussed have so far been only those of non-interacting systems which only
feel a $U(1)$ real space gauge field, though the $k$-space Berry
connection has been allowed to be non-Abelian. The latter constraint
is not an inherent limitation of the theory; perhaps there are interesting
responses to a larger real-space gauge group. $SU(2)$ groups in 4D and 3D have been studied and shown to give topological insulator- and WSM-like responses, respectively\cite{LiSU2LL}. It is thus conceivable that general gauge groups can lead to other topological responses, possibly of phases with emergent fermions such as partons\cite{WenNonAbelianPartons,*SwingleFTI,*Maciejko3DFTI} or composite fermions\cite{JainCompositeFermions}.

As for interactions, it is not immediately clear if there
are sensible real space systems that are well-described by a phase
space theory with only local interactions. However, if there are such
real space systems, then working in phase space could be very useful
because, for example, in the absence of a magnetic field mean field
theory is more accurate due to the higher dimensionality. This advantage
may be mitigated by the fact that our construction requires gauge
fields, however. Alternatively, it is possible that there is a simple
way to directly incorporate the interactions of the real space system
into the phase space theory.

Another interesting question is if there is an extension of our theory
which describes nodal superconductors. Our theory as written requires
$U(1)$ charge conservation; perhaps there is some way to incorporate
spontaneous breaking of this symmetry. Finally, it could also be interesting
to explicitly incorporate other symmetries of the lower-dimensional
system; this could allow a better understanding of gapless symmetry-protected
phases like Dirac semimetals.

\begin{acknowledgments}
DB is supported by the National Science Foundation Graduate Research Fellowship under Grant No. DGE-114747. PH is supported by the David and Lucile Packard Foundation and the U.S. DOE, Office of Basic Energy Sciences, contract DEAC02-76SF00515. SCZ is supported by the National Science Foundation under grant No. DMR-1305677. XLQ is supported by the National Science Foundation through the grant No. DMR-1151786.
\end{acknowledgments}
\bibstyle{apsrev4-1} \bibliography{references}

\appendix

\section{Topological Magnetoelectric Effect}

\label{app:TME} Here we derive explicitly the topological magnetoelectric
effect from our response theory. The topological magnetoelectric effect
is only quantized in gapped systems, so we assume the system is gapped.

The relevant terms are in Eqn. (\ref{eqn:unsimplifiedTME}), which
we rewrite here as
\begin{equation}
j_{3D}^{x}=\frac{-1}{16\pi^{3}}\int d^{3}\bar{x}\text{tr}\left[\epsilon^{IJK}\mathcal{F}_{tI}\mathcal{F}_{JK}\mathcal{F}_{yz}\right]\label{eqn:appUnsimpTME}
\end{equation}
where $I,J,K$ run over $\bar{x},\bar{y},\bar{z}$. We assume that
$\mathcal{F}_{yz}$ is the real-space magnetic field $B_{x}$, so
we can pull it out. For simplicity choose a gauge such that $A_{\bar{i}}=0$
for $i=x,y,z$.

Expanding Eq. (\ref{eqn:appUnsimpTME}) and manipulating some indices
yields \begin{widetext}
\begin{equation}
j_{3D}^{x}=-\frac{B_{x}}{8\pi^{3}}\int d^{3}\bar{x}\epsilon^{IJK}\text{tr}\left[(\partial_{t}a_{I}-\partial_{I}a_{t}+[a_{t},a_{I}])(\partial_{J}a_{K}+a_{J}a_{K})\right]
\end{equation}
\end{widetext}

We first show that several sets of terms in this expansion are zero.
First, notice that
\begin{align}
\int d^{3}\bar{x}\epsilon^{IJK}\partial_{I}a_{t}\partial_{J}a_{K} & =-\int d^{3}\bar{x}\epsilon^{IJK}\partial_{I}\partial_{J}a_{t}a_{K}\\
 & =\int d^{3}\bar{x}\epsilon^{IJK}\partial_{J}a_{t}\partial_{I}a_{K}\\
 & =-\int d^{3}\bar{x}\epsilon^{IJK}\partial_{I}a_{t}\partial_{J}a_{K}
\end{align}
where we integrated by parts twice and then switched the indices $I$
and $J$. Hence
\begin{equation}
\int d^{3}\bar{x}\epsilon^{IJK}\partial_{I}a_{t}\partial_{J}a_{K}=0
\end{equation}
Next consider the terms \begin{widetext}
\begin{align}
\int d^{3}\bar{x}\epsilon^{IJK}\text{tr}[[a_{t},a_{I}]\partial_{J}a_{K}-\partial_{I}a_{t}a_{J}a_{K}] & =\int d^{3}\bar{x}\epsilon^{IJK}\text{tr}[-\partial_{J}(a_{t}a_{I}-a_{I}a_{t})a_{K}-\partial_{I}a_{t}a_{J}a_{K}]\\
 & =\int d^{3}\bar{x}\epsilon^{IJK}\text{tr}[\partial_{I}(a_{t}a_{J}-a_{J}a_{t})a_{K}-\partial_{I}a_{t}a_{J}a_{K}]\\
 & =\int d^{3}\bar{x}\epsilon^{IJK}\text{tr}[(a_{t}\partial_{I}a_{J}-\partial_{I}a_{J}a_{t}-a_{J}\partial_{I}a_{t})a_{K}]\\
 & =-\int d^{3}\bar{x}\epsilon^{IJK}\text{tr}[[a_{t},a_{I}]\partial_{J}a_{K}-\partial_{I}a_{t}a_{J}a_{K}]
\end{align}
\end{widetext} We have integrated by parts, manipulated indices,
and used the cyclic property of the trace. Hence the left-hand side
here is also zero.

Finally, trivial manipulations show that $\int d^{3}\bar{x}\epsilon^{IJK}\text{tr}[[a_{t},a_{I}]a_{J}a_{K}]=0$
as well.

The remaining terms in the expansion are those that do not involve
$a_{t}$:
\begin{widetext}
\begin{align}
j_{3D}^{x} & =\frac{-B_{x}}{8\pi^{3}}\int d^{3}\bar{x}\epsilon^{IJK}\text{tr}\left[\partial_{t}a_{I}(\partial_{J}a_{K}+a_{J}a_{K})\right]\\
 & =\frac{-B_{x}}{8\pi^{3}}\int d^{3}\bar{x}\epsilon^{IJK}\text{tr}\left[\partial_{t}(a_{I}\partial_{J}a_{K}+a_{I}a_{J}a_{K})-a_{I}(\partial_{t}\partial_{J}a_{K}+2\partial_{t}a_{J}a_{K})\right]\\
 & =\frac{-B_{x}}{8\pi^{3}}\int d^{3}\bar{x}\epsilon^{IJK}\text{tr}\left[\partial_{t}(a_{I}\partial_{J}a_{K}+a_{I}a_{J}a_{K})+\partial_{J}a_{I}\partial_{t}a_{K}-2\partial_{t}a_{I}a_{J}a_{K})\right]\\
 & =\frac{-B_{x}}{8\pi^{3}}\int d^{3}\bar{x}\epsilon^{IJK}\text{tr}\left[\partial_{t}(a_{I}\partial_{J}a_{K}+a_{I}a_{J}a_{K})-\partial_{t}a_{I}\partial_{J}a_{K}-\partial_{t}a_{I}a_{J}a_{K}-\frac{1}{3}\partial_{t}(a_{I}a_{J}a_{K})\right]\\
 & =\frac{-B_{x}}{8\pi^{3}}\int d^{3}\bar{x}\epsilon^{IJK}\text{tr}\left[\partial_{t}\left(a_{I}\partial_{J}a_{K}+\frac{2}{3}a_{I}a_{J}a_{K}\right)\right]-j_{3D}^{x}
\end{align}
\end{widetext}
This immediately gives the desired relation (\ref{eqn:TME}).
\end{document}